\documentclass[journal]{IEEEtran}
\UseRawInputEncoding
\usepackage{amsfonts}
\usepackage{bm}
\usepackage{amsmath,amsfonts,amsthm,amssymb}
\usepackage{setspace}
\usepackage{Tabbing}
\usepackage{fancyhdr}
\usepackage{lastpage}
\usepackage{extramarks}
\usepackage{chngpage}
\usepackage{algorithmic}
\usepackage{soul,color}
\usepackage{epsfig}
\usepackage{graphicx,float,wrapfig,subfigure}
\usepackage{dsfont}
\usepackage{longtable}
\usepackage{wrapfig}
\usepackage{stfloats}
\usepackage{cite}
\usepackage{graphicx}
\usepackage{array}
\usepackage{multirow}{\tiny }
\usepackage{multicol}
\usepackage{colortbl}
\usepackage{tabularx}
\usepackage{mdwmath}
\usepackage{mdwtab}
\usepackage{color}
\usepackage{verbatim}
\usepackage{amsmath}
\usepackage{tikz}
\usetikzlibrary{calc}
\usepackage{flowchart}
\usetikzlibrary{shapes.geometric, arrows}
\usepackage{overpic}
\usepackage{epstopdf}
\usepackage{url}
\usepackage[colorlinks,linkcolor=black,anchorcolor=black,citecolor=black,urlcolor=black]{hyperref} 
\usepackage{caption}
\usepackage{verbatim}%多行注释
\usepackage{nomencl}%所用宏包
\makenomenclature%必须加上，放在\begin{document}之前
\renewcommand{\nomgroup}[1]{
\item[\textit{\ifthenelse{\equal{#1}{C}}{C. Parameter}{}\ifthenelse{\equal{#1}{D}}{D. Variable}{}\ifthenelse{\equal{#1}{B}}{B. Set}{}\ifthenelse{\equal{#1}{A}}{A. Abbreviation}{}% add more groups as needed
    }]%

}

\usepackage[justification=centering]{caption} %figures caption centering
\newtheorem{remark}{Remark}

\usepackage{color}
\usepackage{hyperref}
\usepackage[hyphenbreaks]{breakurl}
\makeatletter % 
\let\myorg@bibitem\bibitem
\def\bibitem#1#2\par{%
	\@ifundefined{bibitem@#1}{%
		\myorg@bibitem{#1}#2\par
	}{%
		\begingroup
		\color{\csname bibitem@#1\endcsname}%
		\myorg@bibitem{#1}#2\par
		\endgroup
	}%
}

\makeatother % 
\allowdisplaybreaks

\usepackage{footnote}
\makesavenoteenv{tabular}
\makesavenoteenv{table}

\begin{document}

\title{
%	\vspace{-4mm}
	 District Cooling System Control for Providing Operating Reserve based on Safe Deep Reinforcement Learning
	}

\author{
Peipei~Yu,~\IEEEmembership{Student Member,~IEEE,}
Hongxun~Hui,~\IEEEmembership{Member,~IEEE,}
Hongcai~Zhang,~\IEEEmembership{Member,~IEEE,}
%Wei~Qi,
Ge~Chen,~\IEEEmembership{Student Member,~IEEE,}
and~Yonghua~Song,~\IEEEmembership{Fellow,~IEEE}
\vspace{-6mm}

\thanks{
This paper is funded in part by the Science and Technology Development Fund, Macau SAR (File no. SKL-IOTSC(UM)-2021-2023, 0137/2019/A3, and 0003/2020/AKP), and in part by the National Natural Science Foundation of China under Grant 52007200.  (Corresponding author: \textit{Hongcai Zhang}.)

P. Yu, H. Hui, H. Zhang, G. Chen, and Y. Song are with the State Key Laboratory of Internet of Things for Smart City and Department of Electrical and Computer Engineering, University of Macau, Macao, 999078 China; H. Zhang is also with the Smart City Research Center, Zhuhai UM Science \& Technology Research Institute, Zhuhai, 519031 China (email: hczhang@um.edu.mo). 
	}

}

\maketitle

\begin{abstract} 
	Heating, ventilation, and air conditioning (HVAC) systems are well proved to be capable to provide operating reserve for power systems. As a type of large-capacity and energy-efficient HVAC system (up to 100 MW), district cooling system (DCS) is emerging in modern cities and has huge potential to be regulated as \textcolor{black}{a flexible load}. 
	%DCS很大、很多、有潜力
	However, strategically controlling a DCS to provide flexibility is challenging, because one DCS services multiple buildings with complex thermal dynamics and uncertain cooling demands. Improper control may lead to significant thermal discomfort and even deteriorate the power system's operation security. To address the above issues, we propose a model-free control strategy based on the deep reinforcement learning (DRL) without the requirement of accurate system model and uncertainty distribution.
	%DCS很复杂、很多不确定性, 我们提出了model-free DRL解决了控制问题
	To avoid damaging ``trial \& error" actions that may violate the system's operation security during the training process, we further propose a safe layer combined to the DRL to guarantee the satisfaction of critical constraints, forming a safe-DRL scheme. 
	%避免DRL的training process对电网不安全, 我们进一步提出了safe layer 解决了安全问题
	Moreover, after providing operating reserve, DCS increases power and tries to recover all the buildings' temperature back to set values, which may probably cause an instantaneous peak-power rebound and bring a secondary impact on power systems. Therefore, we design a self-adaption reward function within the proposed safe-DRL scheme to constrain the peak-power effectively.
	%退出控制时, 由于系统想尽快恢复用户设定温度, 会导致功率反弹, safe-DRL框架结合xxx可以很好地解决这个问题
	Numerical studies based on a realistic DCS demonstrate the effectiveness of the proposed methods. 
\end{abstract}
%\vspace{-4mm}
\begin{IEEEkeywords}
District cooling system, operating reserve, model-free control, safe deep reinforcement learning.
\end{IEEEkeywords}

\textcolor{black}{\printnomenclature
}
%变量类型、变量名称、变量解释
%\nomenclature[xxx]{$xxx$}{xxx}
\nomenclature[A]{DCS}{District cooling system}
\nomenclature[A]{HVAC}{Heating, ventilation, and air conditioning}
\nomenclature[A]{DRL}{Deep reinforcement learning}
\nomenclature[A]{MDP}{Markov decision process}
\nomenclature[A]{CMDP}{Constrained Markov decision process}
\nomenclature[A]{AHU}{Air handling unit}
\nomenclature[A]{PI}{Proportional integral}

\nomenclature[D]{$P^\text{ch}_t$}{Electrical power of chillers}
\nomenclature[C]{$P^\text{cap}$}{Power cap during operating reserve}
\nomenclature[D]{$Q^\text{ch}_t$}{Cooling power of chillers}
\nomenclature[D]{$Q^\text{HE}_{i,t}$}{Exchanging heat of building $i$ in the heat exchanger}
\nomenclature[D]{$Q^\text{DCS}_{i,t}$}{Cooling gain from DCS in building $i$}
\nomenclature[D]{$Q^\text{loss}_{i,t}$}{Heat loss in building $i$}
\nomenclature[D]{$\zeta_{i,t}$}{Heat loads from indoor sources in building $i$}

\nomenclature[D]{$m^\text{ch}_t$}{Total water mass flow of chillers}
\nomenclature[D]{$m^\text{dec}_t$}{Water mass flow of the decoupler}
\nomenclature[D]{$m^\text{I}_{i,t}$}{Water mass flow of building $i$ in first water loop}
\nomenclature[D]{$m^\text{II}_{i,t}$}{Water mass flow of building $i$ in second water loop}
\nomenclature[D]{$m^\text{w}_{i,t}$}{Wind mass flow of building $i$ in air loop}
\nomenclature[D]{$\Delta m^\text{I}_{i,t}$}{Regulation of mass flow rate in first water loop}
\nomenclature[C]{$\overline m^\text{I}_{i}$}{Upper limit of mass flow rate in first water loop}
\nomenclature[C]{$\underline m^\text{I}_{i}$}{Lower limit of mass flow rate in first water loop}

\nomenclature[D]{$T^\text{ch,r}_t$}{Return water temperature of chillers}
\nomenclature[D]{$T^\text{dec}_t$}{Water temperature of the decoupler}
\nomenclature[D]{$T^\text{I,r}_{i,t}$}
{Return water temp of building $i$ in first water loop}
\nomenclature[D]{$T^\text{I,s}_{i,t}$}{Supply water temp of building $i$ in first water loop}
\nomenclature[D]{$T^\text{II,r}_{i,t}$}
{Return water temp of building $i$ in second water loop}
\nomenclature[D]{$T^\text{II,s}_{i,t}$}
{Supply water temp of building $i$ in second water loop}
\nomenclature[D]{$\Delta T^\text{mean}_{i,t}$}{Mean temperature difference in heat exchanger}
\nomenclature[D]{$T^\text{A}_{i,t}$}{Indoor temperature of building $i$}
\nomenclature[D]{$T^\text{out}_{i,t}$}{Ambient temperature}
\nomenclature[D]{$T^\text{w}_{i,t}$}{Cooling wind temperature of building $i$}
\nomenclature[D]{$\Delta T_{i,t}$}{Temperature deviation from set value in building $i$}

\nomenclature[D]{$\bm{s}_t$}{State of the environment}
\nomenclature[D]{$\bm{s}_t$}{Action of the agent}
\nomenclature[D]{$\tau$}{Complete trajectory of the policy}
\nomenclature[D]{$\pi$}{Policy of the agent}
\nomenclature[D]{$r_t$}{Reward function of the action}
\nomenclature[D]{$G_t$}{Expected return of the action}
\nomenclature[D]{$J_\pi$}{Expected return of the policy}
\nomenclature[C]{$\gamma$}{Discounted factor}
\nomenclature[D]{$Q^\pi$}{Action-value function}
\nomenclature[D]{$\theta^Q,\theta^{Q'}$}{Parameters of $Q$ and $Q'$ network}
\nomenclature[D]{$\theta^{\pi},\theta^{\pi'}$}{Parameters of $\pi$ and $\pi'$ network}
\nomenclature[D]{$\mu_t,\upsilon_t$}{Correction coefficient for regulation}

\nomenclature[C]{$T^\text{ch,s}$}{Supply water temperature of chillers}
\nomenclature[C]{$\text{COP}$}{Chillers' coefficient of performance}
\nomenclature[C]{$c^\text{w}$}{Specific heat capacity of the water}
\nomenclature[C]{$c^\text{A}$}{Specific heat capacity of the air}
\nomenclature[C]{$\rho^\text{A}$}{Density of the air}
\nomenclature[C]{$V_i$}{Space volume of building $i$}
\nomenclature[C]{$\eta^{\text{I}}_i$}{Transfer efficiency of first water loop}
\nomenclature[C]{$\eta^{\text{II}}_i$}{Transfer efficiency of second water loop}
\nomenclature[C]{$k^{\text{HE}}_i$}{Transfer coefficient of heat exchanger in building $i$}
\nomenclature[C]{$F^{\text{HE}}_i$}{Surface area of heat exchanger in building $i$}
\nomenclature[C]{$\alpha_i$}{Mixing proportion of the fresh air in building $i$}
\nomenclature[C]{$U^\text{O-A}_i$}{Heat transfer coefficient of building $i$}
\nomenclature[C]{$A^\text{S}_i$}{Surface area of building $i$}

\nomenclature[B]{$\mathcal{I}$}{Set of served buildings}
\nomenclature[B]{$\mathcal{T}$}{Set of time slots}
\nomenclature[B]{$\mathcal{S}$}{Set of agent's states}
\nomenclature[B]{$\mathcal{A}$}{Set of agent's actions}
\nomenclature[B]{$\mathcal{K}$}{Set of mini-batch sampled data}

\section{Introduction}
\subsection{Background}
\IEEEPARstart{T}{he} \textcolor{black}{increasing intermittent renewable energy resources bring more uncertainties to the generation-side, and scale up the demands for operating reserve services in modern power systems \cite{RENS2020}}. %改了参考文献
Traditionally, the service is majorly provided by thermal or gas generating units, which are carbon-intensive and being phased out \cite{reserve2015}. With the development of Internet of Things technologies, active control of demand-side resources has emerged as an alternative solution to provide operating reserve by curtailing or transferring power consumption \cite{SIANO2014}. The \textit{heating, ventilation, and air conditioning} (HVAC) system is an ideal resource, because it can shift its power consumption flexibly while assuring the comfortable indoor temperature by utilizing the building's inherent thermal inertia \cite{Luning2021}. Besides, HVAC has large regulation capacity as it accounts for over 40\% of power consumption in modern cities \cite{Cai2019}.

%\textcolor{black}{problem statement -- what problem are we studying and what are the controlling objectives? challenges: complexity and uncertainty. and What if your control is not good enough. }, and thus has large regulation capacity 

\begin{figure*}
	\centering
	%\vspace{-6mm}
	\includegraphics[width=1.9\columnwidth]{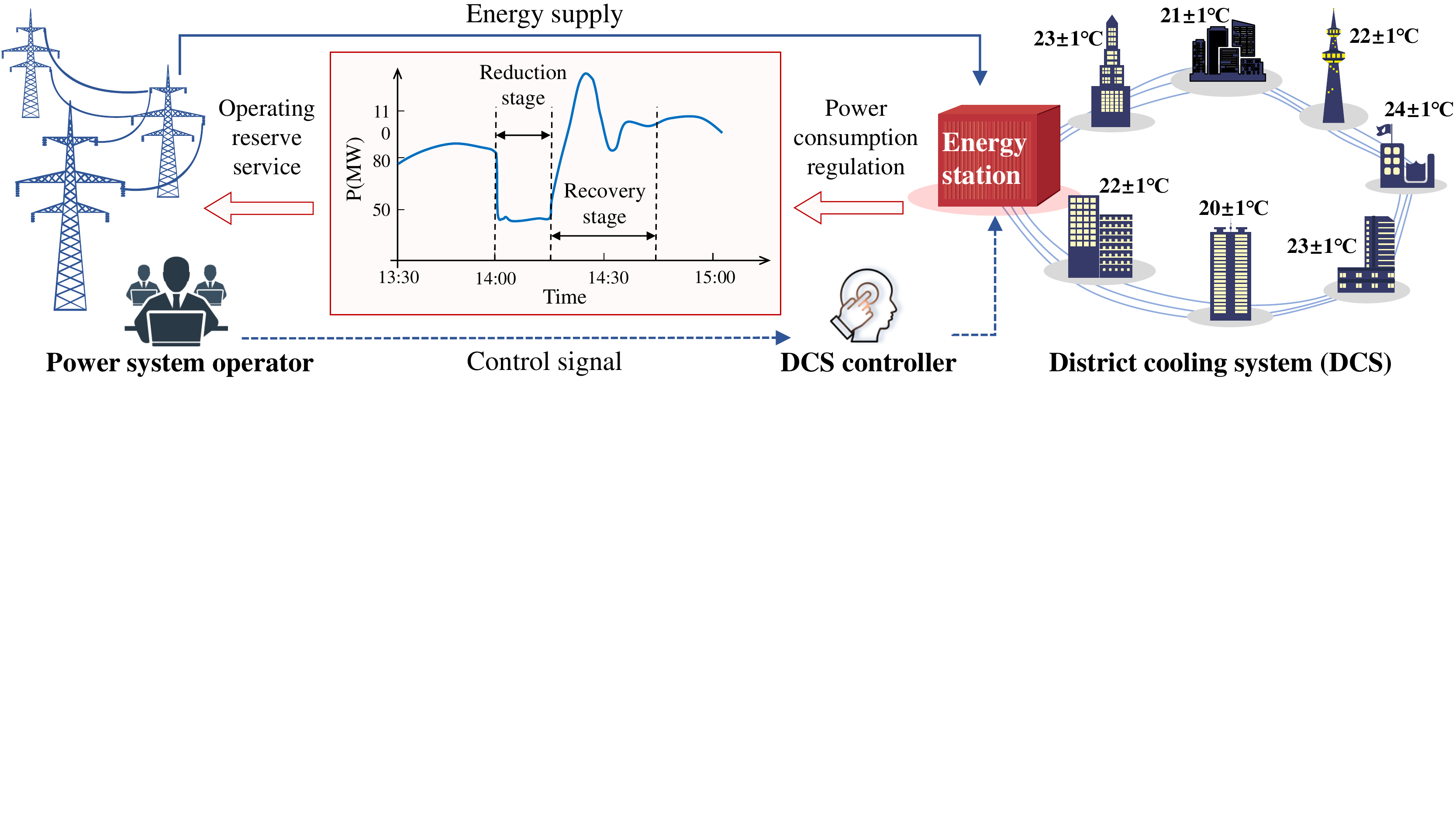}
	\vspace{-50mm}
	\caption{The supply methods and scale of the DCS.}
	\vspace{-4mm}
	\label{dcs_introduction}
\end{figure*}

Compared with a common household HVAC system, the \textit{district cooling system} (DCS) is one type of HVAC with larger capacity and higher efficiency, and thus DCS is emerging and being developed in many cities \cite{Sven2017}. As shown in Fig.~\ref{dcs_introduction}, DCS is composed of one energy station and some pipelines to produce chilled water for multiple buildings \cite{Tou2019}. Generally, one DCS's capacity can be up to 100 MW, which is more than 10,000 times of a household HVAC \cite{HuQinran2020}. \textcolor{black}{Therefore, DCS has huge regulation potential to provide operating reserve, which, however, is only studied by few published papers. For example, Lo et al. \cite{lo2016ice} use least squares regression to optimize the day-ahead power dispatch for a large cooling system to perform demand response. Cox et al. \cite{Tou2019} and Chen et al. \cite{ChenGe2021} design day-ahead power scheduling strategies for DCS to minimize electricity costs with time-of-use pricing. Tang et al. \cite{tang2018optimal} propose a direct load control strategy for a centralized AC system for requests of smart grids. The chillers are assumed to be operated in the on-off mode.
All of the aforementioned studies focus on the day-ahead or hour-ahead control while ignore the real-time uncertainties in cooling demands. Besides, the operation mode of chillers is usually continuous so that assuming it to be on-off mode may not fully utilize DCS's regulation capacity.}
To fill this research gap, this study focuses on the real-time control of a DCS to provide operating reserve subject to the comfortable temperature constraint in each building. 
\textcolor{black}{In most electricity markets, the start time for resources (i.e, DCS) to provide operating reserve is uncertain, while the time interval for operating reserve is fixed (e.g., 10 minutes in PJM \cite{PJMrule01}, 15$\sim$30 minutes in China \cite{2017Jiangsu}). }
As illustrated by the load curve in Fig.~\ref{dcs_introduction}, there are two control stages for a DCS to provide operating reserve:
\begin{enumerate}
    \item \textit{In the power reduction stage}, the controller cuts down the DCS operating power following the instruction from the power system operator. In the meantime, it also tries to fulfill the temperature requests of heterogeneous buildings, when the cooling supply from DCS gets decreased as a result of power reduction. 
    \item \textit{In the power recovery stage}, the DCS stops providing reserve and begins to restore all the buildings' indoor temperature back to set values by increasing its cooling supply. During this stage, the DCS shall recover its power consumption smoothly to avoid the peak-power rebound that may cause a secondary impact on the power system, which has just returned to the stable state.
\end{enumerate}

The above two-stage control of DCS is quite challenging because of both the system complexity and cooling demand uncertainty, detailed as follows: 

\textit{Complexity}: The power consumption of a DCS is usually adjusted automatically according to its operation state, so that we cannot control the power directly to provide operating reserve like other demand-side resources \cite{Simon2021}. 
\textcolor{black}{Instead, the mass flow in pipelines is regulated to adjust the DCS's operating power indirectly \cite{zheng2021incentive}, which can achieve faster-response effects than regulating the set temperature \cite{lu2008redbook}.}
However, describing the relationship between the mass flow and operating power needs an accurate thermal dynamic model. 
This is challenging because the thermal dynamics of a DCS, including cooling power generation, transportation and consumption, is usually quite complex \cite{Sven2017}.
\textcolor{black}{Furthermore, a DCS is a large networked system whose dynamic parameters are hard and expensive to measure. Obtaining an accurate model that is completely in sync with the real system is nontrivial in practice \cite{DCSreview2017}.}
Therefore, traditional model-based control methods for HVAC systems are hard to be used in DCS \cite{Ding2019}. 

\textit{Uncertainty}: The DCS's power consumption and buildings' indoor temperatures are related to the ambient temperature and indoor human behaviors \cite{yin2016quantifying}. Higher ambient temperature and more indoor human activities call for more cooling supply and higher power consumption. \textcolor{black}{However, the indoor human behaviors are stochastic and hard to accurately predict \cite{Cai2019}. The ambient temperature can bring different influence to heterogeneous buildings.} As a result, it is nontrivial to control a DCS subject to heterogeneous indoor temperature constraints in multiple buildings, especially when the DCS power consumption is cut down to provide operating reserve.

\subsection{Literature Reviews}
\textcolor{black}{In recent years, some researchers adopt model predictive control (MPC) method for regulating DCS to achieve the cost reduction in energy systems \cite{04MPC2017}. However, MPC usually requires a reliable dynamic model of the system, which is often unavailable in practice \cite{05MPC2020}. Furthermore, when the system model is too complex, MPC can be quite computationally expensive and may fail to work in real-time control scenarios.} Mixed-integer linear programming (MILP) has also been used to the complex system scheduling problems \cite{03MILP2020,04MILP2021}. Unfortunately, at each step, optimization methods need to recalculate from the beginning, resulting in too large calculation cost to be applied in real-time control. Moreover, the execution time of MILP increases exponentially according to the problem dimensions and cannot solve complex issues.
Another commonly used control method is heuristic algorithm. including genetic algorithm (GA), particle swarm optimization (PSO), ants colony optimization (ACO), etc. Stoppato et al. \cite{PSO2016model} combine heuristic algorithms to obtain system's optimal operation, while the convergence of heuristic algorithms cannot be proved mathematically and is less robust.

\textcolor{black}{
Compared with the aforementioned control methods, deep reinforcement learning (DRL) has become increasingly popular to handle model-free and high dimensional decision-making problems\cite{mnih2015human}. DRL has been proved to be more robust with stable convergence results to effectively handle uncertainties of systems through the prediction in neural networks.}
Some researchers have adopted DRL to control traditional HVACs. For instance, Du et al. \cite{YanDu2021} use DRL to control residential HVACs as to respond dynamic electricity prices. Xu et al. \cite{YanXu2020} adopt DRL to schedule home energy consumption considering uncertain PV generation. Liang et al. \cite{LiangYu2021} present a DRL-based control strategy to minimize both the HVAC's energy consumption and user's thermal discomfort. Ruelens et al. \cite{Frederik2019} propose a DRL-based direct control method of HVAC to provide ancillary services. Zhang et al. \cite{Xiangyu2021} utilize DRL for cost-effective control of a HVAC in commercial buildings. However, to the best of our knowledge, published papers have not studied DRL-based control for DCS to provide operating reserve for power systems.

Generally, a DRL-based controller has to be trained through lots of ``trail-and-errors" before being intelligent \cite{DDPG2016}. 
\textcolor{black}{It means some ``bad" decisions may be made during the training process, part of which may cause constraint violations. 
However, in power systems, some critical constraint violations can cause damaging results \cite{DOBBE2020}. For example, if a DCS fails to provide sufficient operating reserve as it promised to the system operator, the power system may face the stability problem.} 
\textcolor{black}{
To address this challenge from the constraint safety in power systems, the safe-DRL framework is needed to ensure the satisfaction of critical constraints during the training process. Considering that safe-DRL is an emerging concept in the application in power systems, there are little research to combine HVAC or DCS with safe-DRL. 
Some published studies propose safe-DRL frameworks for voltage control problems \cite{01voltage2020,02voltage2020,03voltage2019,04voltage2020}, emergency load-shedding control problems \cite{01emerg2021,02emerg2020,03emerg2019} and demand-side resource scheduling problems (e.g., EV, building's equipment) \cite{01schedule2021,02EV2019,03HVAC2019}.
However, DCS is different from EVs and household HVACs due to its complex thermal dynamic process and uncertainties. Thus, the above methods can not be adopted directly in this paper.}

\subsection{Contributions}
In this paper, we propose a safe-DRL control strategy for DCS to provide operating reserve while satisfying major critical constraints. This paper advances the relevant published literature in the following aspects:

\begin{enumerate}
	\item \textcolor{black}{The DCS control problem is developed as a Markov Decision Process (MDP) mathematically to provide operating reserve. The designed reward function of DRL aims to achieve minimum impacts on buildings' thermal comfort when providing the required operating reserve. Besides, the proposed iteration algorithm does not need the accurate system model of DCS nor the distribution of uncertainties, which can address challenges from both the system complexity and uncertainty.}
	\item A novel safe-DRL framework is proposed for constraint assurance, where a safe layer is designed on the top of traditional DRL algorithm. The proposed safe layer avoids critical power constraint violations to protect power systems from undesirable ``trial-and-errors", through fine tuning possible unsafe control signals into safe ones during the training process.
	\item A self-adaption target method is proposed and designed as the reward function in the safe-DRL framework during the power recovery stage. The proposed method can effectively achieve the smooth power recovery and avoid peak-power rebound that probably brings secondary impacts to power systems.
\end{enumerate}

Besides, numerical studies verify the effectiveness of our proposed strategy, based on a real-world DCS. The analysis shows DCSs are qualified to provide operating reserve with mild impacts on buildings' indoor thermal comforts, subject to the critical power constraints.

The rest is organized as follows. Section \ref{dcs_system} introduces the physical architecture and control logic of DCS. Section \ref{dcs_control} proposes the safe-DRL framework. Numerical studies are carried out in Section \ref{dcs_case}. Section \ref{dcs_conclusion} concludes this paper.

\section{Modelling of the DCS}\label{dcs_system}

\textcolor{black}{This section establishes the DCS model as the simulated environment to interact with the proposed DRL agent. Noted that the only information received by the agent is the feedback from the established environment, while not the details about the accurate DCS model.}

\subsection{DCS Framework}
\begin{figure*}
	\centering
	\includegraphics[width=1.8\columnwidth]{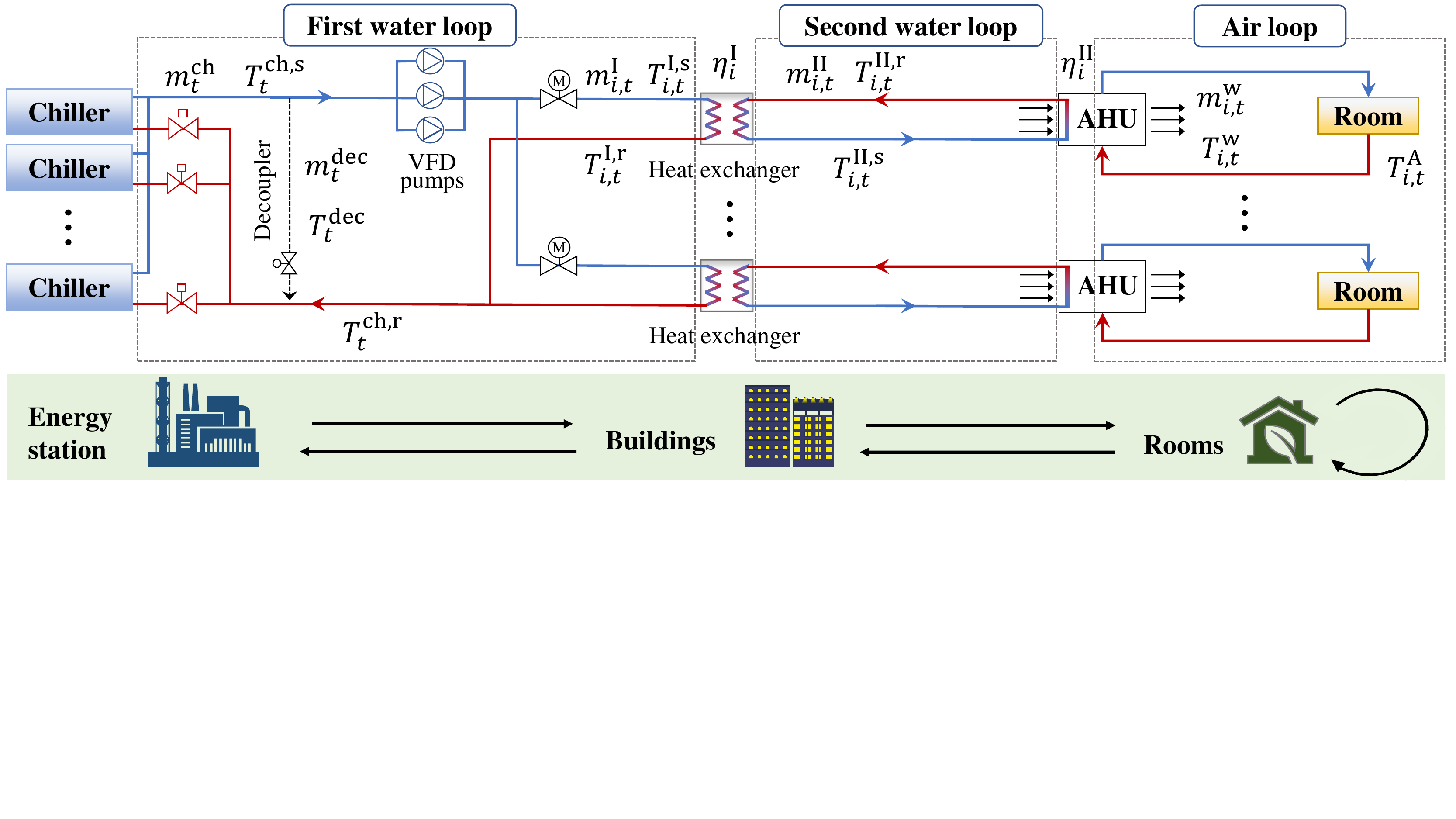}
	\vspace{-38mm}
	\caption{\textcolor{black}{Schematic diagram of a DCS.}}
	\label{fig_dcs_structure}
	\vspace{-4mm} 
\end{figure*}

The schematic diagram of a DCS is shown in Fig.~\ref{fig_dcs_structure}, in which blue lines represent the chilled water (or cooling wind) to supply thermal energies for buildings; red lines are the returned warm water (or warm wind). Its heat transmission process includes three isolated loops:

\textcolor{black}{In the first water loop, chillers produce chilled water with a set temperature $T^\text{ch,s}$, which is transported through pipelines to distributed buildings to supply cooling demands. 
The total mass flow $m^\text{ch}_t$ is separated to different buildings by their independent two-port valves, which determine each building's own mass flow rate $m^\text{I}_{i,t}$ \cite{zheng2021distributed}. 
After the heat exchange process, the chilled water in pipelines becomes warm with temperature $T^\text{ch,r}_{t}$ and then is pumped back to chillers.}
The decoupler between the supply and return water balances pressure when the mass flow rate changes.

\textcolor{black}{In the second water loop (i.e., water cycle in buildings), the water temperature $T^\text{II,s}_{i,t}$ in buildings is cooled down by the chilled water in the first water loop through heat exchangers. 
Then the cool water transfers its thermal energy to the air in Air Handle Units (AHUs) to form cooling winds.}
The temperature of return water $T^\text{II,r}_{i,t}$ reflects fluctuating cooling demands in buildings and further influences chillers' power consumption automatically.

\textcolor{black}{In the air loop, AHUs blow cooling winds with the temperature $T^\text{w}_{i,t}$ into each room, which can further influence the indoor temperature $T^\text{A}_{i,t}$ and refresh the indoor air. }

Note that the aforementioned three loops are mutually independent while interactional. Specifically, the total power consumption of a DCS majorly comes from chillers in the first water loop, whose operations are automatically and indirectly adjusted based on the buildings' cooling demands in the third loop. Therefore, it is significant for the DCS control to find the relationship between these loops.

\subsection{Modelling of Key Components}
\subsubsection{Chillers}
Chillers consume most electricity in DCS. Their power consumption can be calculated based on the energy and mass balance, as follows: 
\begin{align}
	&P^{\text{ch}}_t=\frac{Q^{\text{ch}}_t}{\text{COP}},
	\quad \forall t,\label{eqn_ch_1}
\end{align}
\textcolor{black}{where $P^{\text{ch}}_t$ is chillers' electrical power at time $t$, in kW; $Q^{\text{ch}}_t$ is the cooling power, in kW;} $\text{COP}$ denotes chiller's coefficient of performance. Generally, $Q^{\text{ch}}_t$ is determined by chillers' varying return water temperature $T^{\text{ch,r}}_t$, in $^\circ$C, and instantaneous mass flow rate $m^{\text{ch}}_t$, in kg/s, as follows:
\begin{align}
	&Q^{\text{ch}}_t=m^{\text{ch}}_tc^{\text{w}} (T^{\text{ch,r}}_t-T^{\text{ch,s}}),
	\quad \forall t,\label{eqn_ch_2}
\end{align}
where $c^{\text{w}}$ is the specific heat capacity of water, in kJ/(kg$\cdot^\circ$C). The set temperature of supply chilled water is represented by $T^{\text{ch,s}}$, \textcolor{black}{which is usually a designed constant \cite{Beijing2004}}. 
Therefore, controlling the mass flow rate $m^{\text{ch}}_t$ can influence the electrical power $P^{\text{ch}}_t$ effectively. Further, we can rewrite $T^{\text{ch,r}}_t$ and $m^{\text{ch}}_t$ according to the mass balance as:
\begin{align}
	&m^{\text{ch}}_t=m^{\text{dec}}_t+\sum\nolimits_{i \in \mathcal{I}}m^{\text{I}}_{i,t},
	\quad \forall t,\label{eqn_ch_3}\\
	&T^{\text{ch,r}}_t=\frac{m^{\text{dec}}_tc^{\text{w}}T^{\text{dec}}_t+\sum_{i \in \mathcal{I}}{m^{\text{I}}_{i,t}c^{\text{w}}T^\text{I,r}_{i,t} }}{m^{\text{ch}}_tc^{\text{w}}},
	\quad \forall t,\label{eqn_ch_4}
\end{align}
where set $\mathcal{I}$ denotes the set of terminal buildings; $m^\text{I}_{i,t}$ and $T^\text{I,r}_{i,t}$ are each buildings' mass flow rate and return water temperature in first water loop, respectively; 
\textcolor{black}{$m^{\text{dec}}_t$ and $T^{\text{dec}}_t$ are the mass flow rate and return water temperature of the decoupler, respectively. Eqs. (\ref{eqn_ch_3})-(\ref{eqn_ch_4}) show the mass flow and energy balances between chillers and buildings. }

\subsubsection{Heat exchangers}
\textcolor{black}{Heat exchangers transfer cooling supply from the first water loop to the second water loop.} \textcolor{black}{Considering the heat loss in pipelines, each building's actual supply water temperature can be calculated by \cite{zheng2021dynamic}:}
\begin{align}
    &\textcolor{black}{T_{i,t}^\text{I,s}=T_{t}^\text{out}+\eta^\text{pipe}(T^\text{ch,s}-T_{t}^\text{out})
    ,\quad\forall i\in\mathcal{I},\forall t,}
\end{align}
\textcolor{black}{where $\eta^\text{pipe}$ is the heat transfer coefficient of supply pipelines; $T_{t}^\text{out}$ is the ambient temperature; $T_{i,t}^\text{I,s}$ is the supply chilled water temperature for building $i$.} Further, the corresponding exchanging heat in building $i$, $Q^\text{HE}_{i,t}$ in kW, can be given by:
\begin{align}
	Q^\text{HE}_{i,t}
	&= m^{\text{II}}_{i,t}c^{\text{w}}(T^{\text{II,r}}_{i,t}-T^{\text{II,s}}_{i,t})
	\notag\\
	&=  \eta^{\text{I}}_i m^{\text{I}}_{i,t}c^{\text{w}}(T^{\text{I,r}}_{i,t}-T_{i,t}^\text{I,s}),\quad\forall i\in\mathcal{I},
	\forall t, \label{eqn_HE_1}
\end{align}
where $\eta^{\text{I}}_i$ indicates the transfer efficiency of first water loop to second water loop; $m^{\text{I}}_{i,t}$ and $m^{\text{II}}_{i,t}$ are the mass flow rate of two sides, respectively. Similarly, $T^{\text{I,r}}_{i,t},T_{i,t}^\text{I,s}$ and $T^{\text{II,r}}_{i,t},T^{\text{II,s}}_{i,t}$ are the return and supply water temperature of each side, respectively. In addition, $Q^\text{HE}_{i,t}$ is determined by the performance of the heat exchanger, which can be calculated by \cite{Marshall1983}:
\begin{align}
	&\frac{Q^\text{HE}_{i,t}}{k^{\text{HE}}_i}=
	\int_{0}^{F^{\text{HE}}}\Delta T_{i,t}\mathrm{d}F_i \approx F_i^{\text{HE}}\Delta T^{\text{mean}}_{i,t},
	\quad\forall i\in\mathcal{I},
	\forall t,\label{eqn_HE_3}
\end{align}
where $k^{\text{HE}}_i$ is heat exchangers' transfer coefficient, in kW/(m$^{\text{2}}\cdot^\circ$C); $F_i^{\text{HE}}$ is the surface area, in m$^{\text{2}}$; Symbol $\Delta T^{\text{mean}}_{i,t}$ is defined as the mean difference between the water's temperature of two sides, which is a function formulated as:
\begin{align}
	&\Delta T^{\text{mean}}_{i,t}=\frac{(T^{\text{II,r}}_{i,t}-T_{i,t}^\text{I,s})-(T^{\text{II,s}}_{i,t}-T^{\text{I,r}}_{i,t})}{\ln((T^{\text{II,r}}_{i,t}-T_{i,t}^\text{I,s})/(T^{\text{II,s}}_{i,t}-T^{\text{I,r}}_{i,t}))},
	\quad\forall i\in\mathcal{I}
	,\forall t.\label{eqn_HE_4}
\end{align}
The above Eqs. (\ref{eqn_HE_1})-(\ref{eqn_HE_4}) determine the dynamic exchanging heat in each building $i$ between the first and second water loops.

\subsubsection{Buildings}
AHU transfers the heat from the second water loop to the third air loop by blowing cooling wind whose energy balance is give as:
\begin{align}
	& m^{\text{w}}_{i,t}c^{\text{A}}(T^{\text{A}}_{i,t}-T^{\text{w}}_{i,t})
	= \eta^{\text{II}}_i m^{\text{II}}_{i,t}c^{\text{w}}(T^{\text{II,r}}_{i,t}-T^{\text{II,s}}_{i,t}),
	\forall i\in\mathcal{I},\forall t, \label{eqn_room_1}\\
	&T^{\text{w}}_{i,t}=\frac{1}{2}(1-\alpha_i)(T^{\text{II,s}}_{i,t}+T^{\text{II,r}}_{i,t})+\alpha_i T^{\text{out}}_t,
	\quad\forall i\in\mathcal{I}
	,\forall t,\label{eqn_room_2}
\end{align}
where $c^{\text{A}}, m^{\text{w}}_{i,t}$ are air's specific heat capacity and wind's mass flow rate; $\eta^{\text{II}}_i$ is the exchanging heat efficiency of sencond water loop to AHU; $T^{\text{A}}_{i,t}, T^{\text{out}}$ are the indoor and ambient temperature, respectively; $T^{\text{w}}_{i,t}$ represents the temperature of the cooling air out from AHU, mixing the outdoor fresh air with proportion $\alpha_i$. Then the indoor thermal dynamic is described as \cite{Hui2017}:
\begin{align}
	&c^{\text{A}}\rho^{\text{A}}V_i\frac{\mathrm{d}T^{\text{A}}_{i,t}}{\mathrm{d}t}=Q^{\text{loss}}_{i,t}-Q^{\text{DCS}}_{i,t},
	\quad\forall i\in\mathcal{I},\forall t,\label{eqn_room_3}
\end{align}
where $\rho^{\text{A}}$ is the density of the air, in kg/m$^{\text{3}}$; $V_i$ is the space volume of the $i$th building, in m$^{\text{3}}$; $Q^{\text{loss}}_{i,t}$ is the $i$th building's heat loss because of its heat exchange with the ambient environment and $Q^{\text{DCS}}_{i,t}$ is its cooling gain from DCS, which are given as:
\begin{align}
	&Q^{\text{DCS}}_{i,t}=m^{\text{w}}_{i,t}c^{\text{A}}(T^{\text{A}}_{i,t}-T^{\text{w}}_{i,t}),\quad\forall i\in\mathcal{I},\forall t,\label{eqn_room_4}\\
	&Q^{\text{loss}}_{i,t}=U^{\text{O-A}}_{i}A^{\text{S}}_{i}(T^{\text{out}}_t-T^{\text{A}}_{i,t})+\zeta_{i,t}
	,\quad\forall i\in\mathcal{I},\forall t,\label{eqn_room_5}
\end{align}
where $U^{\text{O-A}}_{i}$ is the heat transfer coefficient, in kW/(m$^{\text{2}}\cdot^\circ$C); $A^{\text{S}}_{i}$ is the surface area of the $i$th building, in m$^{\text{2}}$; $\zeta_{i,t}$ is the heat load from indoor sources (e.g., stochastic human behaviors and electric equipment), in kW.
\textcolor{black}{Eqs. (\ref{eqn_room_3})-(\ref{eqn_room_5}) give the temperature dynamic in buildings. If $\Delta T_{i,t}^\text{A}$ is used to represent the temperature fluctuation in a stable operating state ($T_{i,t_0}^\text{A},m^\text{w}_{i,t_0},T^\text{w}_{i,t_0}$), then the temperature dynamic's first-order Taylor series expression can be expressed as:
\begin{align}
	c^{\text{A}}\rho^{\text{A}}V_i\frac{\mathrm{d}\Delta T^{\text{A}}_{i,t}}{\mathrm{d}t}
	&=-(m^{\text{w}}_{i,t_0}c^{\text{A}}+U^{\text{O-A}}_{i}A^{\text{S}}_{i})\Delta T^{\text{A}}_{i,t}
	+m^{\text{w}}_{i,t_0}c^{\text{A}}\Delta T^\text{w}_{i,t}\notag\\
	&+c^{\text{A}}(T^{\text{w}}_{i,t_0}-T^{\text{A}}_{i,t_0})\Delta m^{\text{w}}_{i,t} \notag
	+U^{\text{O-A}}_{i}A^{\text{S}}_{i}\Delta T^{\text{out}}_{t}\notag\\
	&+\Delta \zeta_{i,t},
	\quad\forall i\in\mathcal{I},\forall t,\label{eqn_dynamics_5}
\end{align}
where $\Delta T^\text{w}_{i,t}$, $\Delta m^{\text{w}}_{i,t}$, $\Delta T^{\text{out}}_{t}$ and $\Delta \zeta_{i,t}$ are the changes in the corresponding four variables. Then, the Laplace transformation of Eq. (\ref{eqn_dynamics_5}) can be obtained as following:
\begin{align}
    &(K_1s+K_2+K_4)T_i^\text{A}(s)&\notag\\
    =&K_2T_i^\text{w}(s)+K_3m_i^\text{w}(s)+K_4T^\text{out}(s)+\zeta_i(s)
    , \forall i \in\mathcal{I},\label{eqn_dynamic_6}
\end{align}
where $K_1=	c^{\text{A}}\rho^{\text{A}}V_i$, $K_2=m^{\text{w}}_{i,t_0}c^{\text{A}}$, $K_3=c^{\text{A}}(T^{\text{w}}_{i,t_0}-T^{\text{A}}_{i,t_0})$ and $K_4=U^{\text{O-A}}_{i}A^{\text{S}}_{i}$. Therefore, the temperature dynamic in each building is an inertial process with inertia time constant $K_1/(K_2+K_4)$, which is mainly determined by each building's inherent characteristics.}

The above models from Eq. (\ref{eqn_ch_1}) to Eq. (\ref{eqn_dynamic_6}) describe the whole thermal dynamics in a DCS. In summary, a DCS provides cooling supply to multiple buildings through two water loops and one air loop to transmit thermal energies.

\begin{remark}
    The chillers' cooling power is not only determined by the mass flow rate $m^{\text{ch}}_t$ but also the uncertain return water temperature $T^{\text{ch,r}}_t$. The later is further influenced by stochastic ambient temperature $T^{\text{out}}$ and heat load $\zeta_{i}$ of buildings in Eq. (\ref{eqn_room_5}). Besides, the accurate thermal model parameters in three loops are unknown and difficult to obtain in practice, which makes the conventional model-based control strategy infeasible for a DCS. To deal with these challenges, a model-free DRL method is proposed in the following Section \ref{dcs_control}. 
\end{remark}

\begin{figure}
	\centering
	\includegraphics[width=1.33\columnwidth]{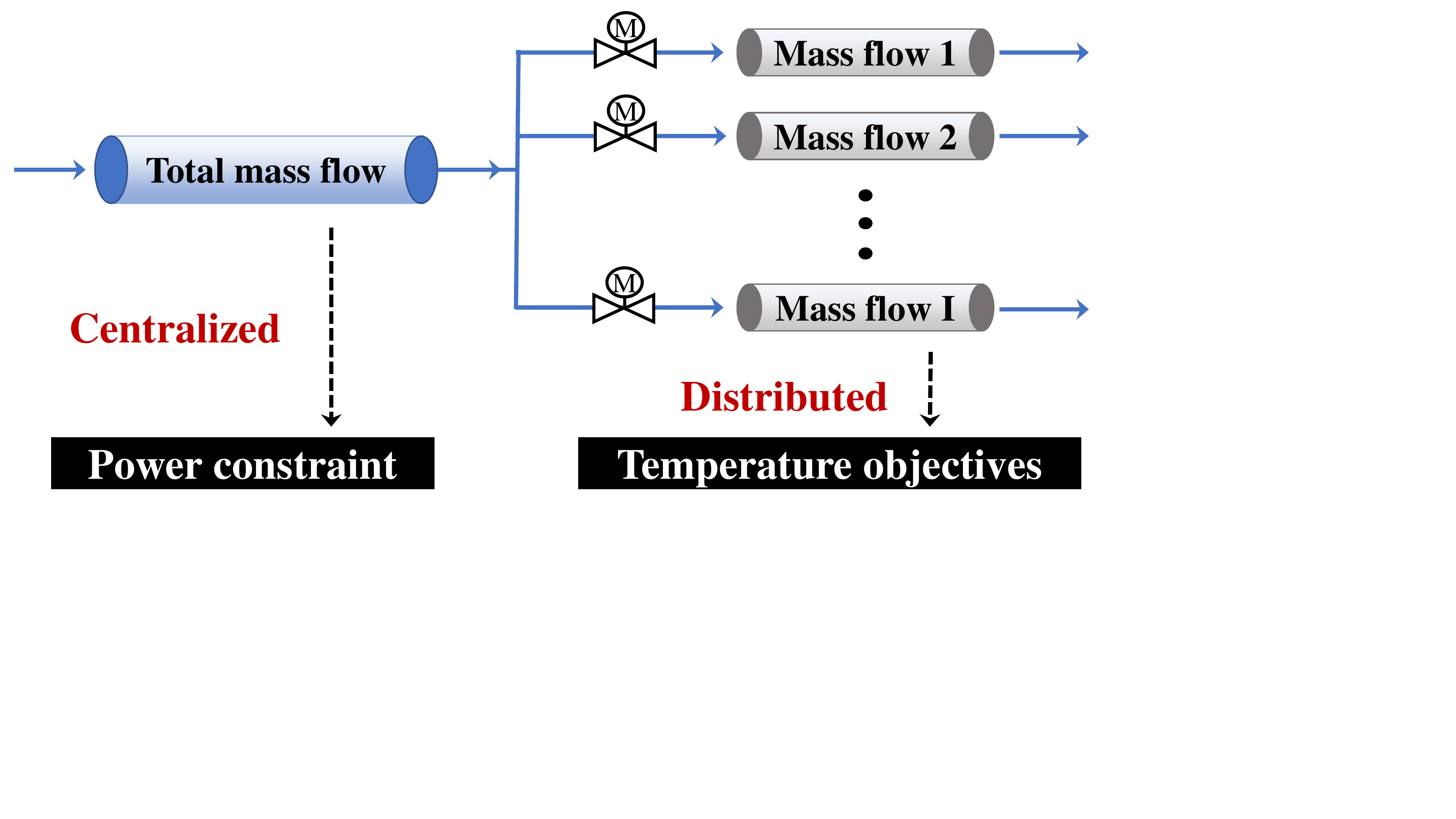}
	\vspace{-28mm}
	\caption{\textcolor{black}{Objectives and constraints during DCS control process.}}
	\label{fig_control_objective}
	\vspace{-4mm}
\end{figure}

\section{DCS Control based on Safe Deep Reinforcement Learning}\label{dcs_control}
\textcolor{black}{As shown in Fig.~\ref{fig_control_objective}, when the DCS receives signals to provide operating reserve, its total mass flow, i.e., the sum of all the buildings' mass flows, should be regulated to satisfy the power constraint. Given the total mass flow, the second problem is to allocate the mass flow among different buildings. An ideal control strategy shall properly allocate these mass flows to make heterogeneous buildings have similar temperature deviations, so that all the buildings' comforts can be guaranteed to the most extent and their thermal inertia can be fully utilized. 
% On contrary, a bad strategy probably results in that some buildings are too hot and some others are too cold, which can not ensure the comfort and not cope with uncertain demands in heterogeneous buildings.
Therefore, the power requirement from the power system and the temperature comfort requirements from buildings should be both considered in the DCS control problem when providing operating reserve.}

\subsection{Formulation of the DCS Control Problem}\label{DDPG_chapter}
\textcolor{black}{The DCS control problem is a typical sequential decision-making problem that can be described as a MDP \cite{2010MDP01}, which aims to minimize temperature impacts on buildings by controlling mass flows continuously. Considering there is a critical power constraint during the control process, the studied problem needs to be formulated as a Constrained Markov Decision Process (CMDP) \cite{altman1999constrained}.}

In the CMDP framework, a centralized smart controller, called \textit{agent}, is designed to send each building signals to control its mass flow rate $m^{\text{I}}_{i,t}$. When a DCS provides operating reserve during the period $\mathcal{T}=[t_{0},t_{1}]$, the DCS is regarded as an \textit{environment} whose real-time operation \textit{state} $\bm{s}_{t}$ at time slot $t\in\mathcal{T}$ is observed by the agent. Then according to the information in $\bm{s}_{t}$, the agent makes 
one decision for DCS to execute \textit{action} $\bm{a}_{t}$ , which means there is a complete trajectory $\tau=\left\{\bm{s}_{t_{0}},\bm{a}_{t_{0}+1},\bm{s}_{t_{0}+1},...,\bm{a}_{t_{1}},\bm{s}_{t_{1}}\right\}$ to describe the control process. 
\textcolor{black}{Here, both the state $\bm{s}_t$ and action $\bm{a}_t$ are multi-dimensional vectors rather than a scalar.}
The probability from the state $\bm{s}_{t}$ to $\bm{s}_{t+1}$ after taking action $\bm{a}_{t}$ is defined by a transition function $P(\bm{s}_{t+1}|\bm{s}_{t},\bm{a}_{t})$, which is not necessary (assumed unknown) in model-free methods.

\begin{figure}[t]
    %\vspace{-2mm}
	\centering
	\includegraphics[width=1.15\columnwidth]{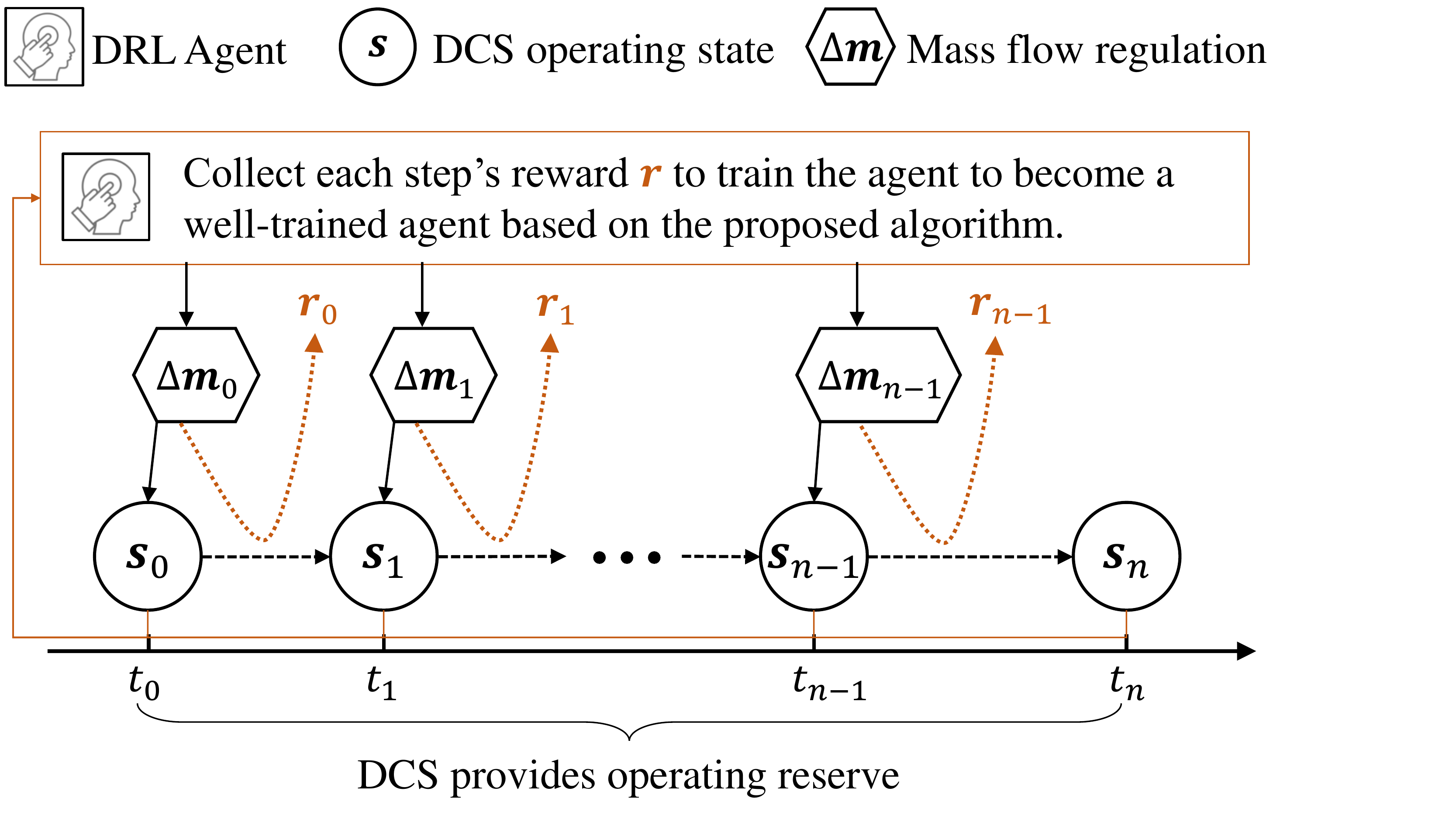}
	\vspace{-6mm}
	\caption{\textcolor{black}{Interactions between the agent and the DCS.}}
	\label{fig_drl_structure}
	\vspace{-4mm}
\end{figure}

In DCS control process, its power consumption and buildings' indoor temperature are main considerations. The temperature deviation $\Delta T_{i,t}=T^{\text{A}}_{i,t}-T^{\text{set}}_{i,t},\forall i\in\mathcal{I},t\in\mathcal{T}$, is defined as the temperature comfort indicator, in which $T^{\text{set}}_{i,t}$ is the set temperature. Thus, the state and the action are defined by: 
\begin{align}
&\bm{s}_{t}=\left[\Delta P_t,m^{\text{I}}_{i,t},T^{\text{I,r}}_{i,t},\Delta T_{i,t} | i\in \mathcal{I}\right]^\intercal \in \mathcal{S},\quad\forall t\in\mathcal{T},\label{eqn_drl_1}\\
&\bm{a}_{t}=\left[\Delta m^{\text{I}}_{1,t},\Delta m^{\text{I}}_{2,t},...,\Delta m^{\text{I}}_{|\mathcal{I}|,t}\right]^\intercal \in \mathcal{A},\quad\forall t\in\mathcal{T},\label{eqn_drl_2}
\end{align}
where $\Delta P_t$ equals to the gap between the actual power $P^\text{ch}_t$ and required power cap $P^\text{cap}$ of power systems. The scale of the state space $\mathcal{S}$ and action space $\mathcal{A}$ are \textcolor{black}{$|\mathcal{S}|=3|\mathcal{I}|+1$} and $|\mathcal{A}|=|\mathcal{I}|$, respectively. 
\textcolor{black}{As shown in Fig.~\ref{fig_drl_structure}, the DRL agent gives its control signal $\Delta \bm{m}$ at each time slot $t$ during operating reserve. According to the past experiences, the agent updates its policy with the proposed algorithm to become more intelligent.}
Because the mass flow can be regulated by valves continuously, the action space is a continuous space and $\Delta m^{\text{I}}_{i,t}$ is a continuous variable. The positive (or negative) $\Delta m^{\text{I}}_{i,t}$ means to increase (or decrease) the mass flow rate, in which there are upper and lower limits $\underline{{m}^{\text{I}}_{i}}$, $\overline{{m}^{\text{I}}_{i}}$ in a real DCS. Thus, the maximum value of action is constrained by $|\Delta m_{i,t}|\leq\overline{{m}^{\text{I}}_{i}}$.

An arbitrary mapping from the state space to the action space $\pi:\mathcal{S}\rightarrow\mathcal{A}$ is called a \textit{policy}. Essentially, the agent's task is to find an \textit{optimal policy} that will be used as a guide for future online controlling. In order to evaluate a policy's performance, $r_{t+1}$ is defined as \textit{reward} for the action $\bm{a}_{t}$ in one step, which is formulated as:
\begin{align}
	&r_{t+1}=-\theta^\text{r}\mathbb{E}_{i\in\mathcal{I}}[|\Delta T_{i,t+1}|] - \sigma^{2}_{i\in\mathcal{I}}[\Delta T_{i,t+1}],~\forall t\in\mathcal{T}.\label{eqn_drl_3}
\end{align}
Eq.~(\ref{eqn_drl_3}) includes two parts: the average and variance of all the buildings' temperature deviation at next time $t+1$. 
\textcolor{black}{The former item $\mathbb{E}_{i\in\mathcal{I}}[|\Delta T_{i,t+1}|]$ is the average temperature of all the buildings' indoor temperature deviations from their corresponding set values. A smaller average value means a less temperature influences to buildings.
The later item $\sigma^2_{i\in\mathcal{I}}[\Delta T_{i,t+1}]$ is the variance of all the buildings' temperature deviations, where a smaller variance means less difference of the influences among different buildings.
Parameter $\theta^\text{r}$ is the weight factor to determine the importance of the two parts. }

\textcolor{black}{Further, compared with the immediate reward $r_t$, the \textit{return} $G_t$ is defined as the accumulated reward in the future, which considers not only the immediate reward but also the expected influence to future rewards caused by the current action. The total discounted reward at time slot $t$ is expressed as:}
\begin{align}
    &G_t=r_{t+1}+\gamma r_{t+2}+...=\sum\nolimits_{\tau=0}^{t_1-t}\gamma^{\tau} r_{t+\tau+1},~\forall t\in \mathcal{T},
\end{align}
\textcolor{black}{where $\gamma \in [0,1]$ is a discount factor to represent the weight of the influence to future rewards \cite{alagoz2010markov}. 
For instance, when $\gamma=1$, the agent considers the immediate and future rewards with the same importance. By contrary, when $\gamma=0$, the agent only considers the current reward and $G_t=r_t$}
Then, an \textit{action-value function} $Q^{\pi}(\bm{s}_{t},\bm{a}_{t})$ is defined as the expected return from state $\bm{s}_{t}$, taking action $\bm{a}_{t}$ and following policy $\pi$:
\begin{align}
	&Q^{\pi}(\bm{s}_{t},\bm{a}_{t})=\mathbb{E}_{\pi}[G_t|\bm{s}_{t},\bm{a}_{t}],\quad\forall t\in\mathcal{T},\label{eqn_drl_4}
\end{align}
where the \textit{optimal action-value function} $Q^*(\bm{s}_{t},\bm{a}_{t})$ means the maximum action-value over all policies $\max_{\pi}Q^{\pi}(\bm{s}_{t},\bm{a}_{t})$. According to the theorem in MDP \cite{DDPG2016}, optimal policy $\pi^*$ is defined to satisfy $Q^{\pi^*}(\bm{s}_{t},\bm{a}_{t})=Q^*(\bm{s}_{t},\bm{a}_{t}),\forall t$. Therefore, the agent's objective is to maximize the expected return $J^{\pi}$:
\begin{align}
	&\max_{\pi} J^{\pi}
	=\mathbb{E}_{s_{t}\sim\mathcal{S},a_t\sim\pi}[G_t]
	=\mathbb{E}_{s_{t}\sim\mathcal{S}}[Q^{\pi}(\bm{s}_{t},\pi(\bm{s}_{t}))]
	.\label{eqn_drl_5}
\end{align}

However, different with the conventional policy optimization problem, there is a critical power constraint for a DCS during the power reduction stage and formulated as:
\begin{align}
	&P^{\text{ch}}_t\leq P^{\text{cap}},\quad\forall t\in\mathcal{T},\label{eqn_drl_6}
\end{align}
\textcolor{black}{where $P^{\text{cap}}$ is the required power cap from the power system operator to constrain DCS's operating power\footnote{\textcolor{black}{Power cap is determined by the regulation capacity that the DCS offered to the electrical market ahead of one day \cite{market2018}.}}}. If it is violated, the DCS may be heavily penalized by the power system operator. Thus, Eq. (\ref{eqn_drl_6}) turns the DCS control problem from a traditional MDP into a CMDP.
% in which we use the policy gradient to update policy $\pi\rightarrow\pi^*$. 

\subsection{Policy Gradient Algorithm}\label{DDPG}

\begin{figure}
    %\vspace{-6mm}
	\centering
	\includegraphics[width=1.1\columnwidth]{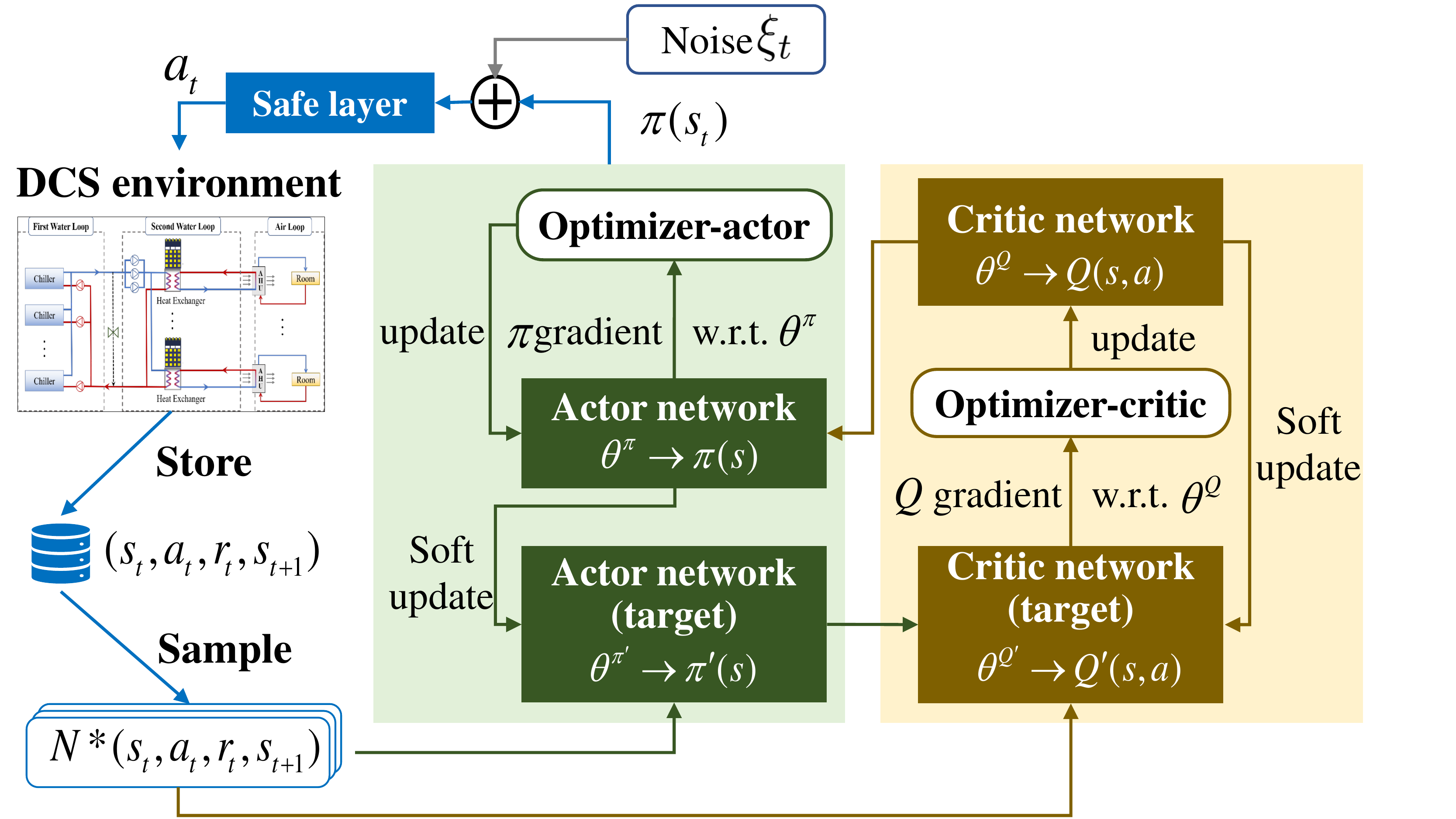}
	\caption{Scheme of the safe-DRL algorithm.}
	\label{DDPG_logic}
	\vspace{-4mm}
\end{figure}

To solve the optimal policy $\pi^*$ in Eq. (\ref{eqn_drl_5}), a safe-DRL algorithm is proposed as shown in Fig.~\ref{DDPG_logic}, which combines the actor-critic framework and deep Q-learning. Two neural networks are adopted to represent the action-value function $Q$ and policy $\pi$, with parameters $\theta^{Q}$, $\theta^{\pi}$, respectively. The network to approximate $Q$ value is called \textit{critic network}, and another one that outputs actions is called \textit{actor network}. In Fig.~\ref{DDPG_logic}, the agent firstly interacts with the DCS environment to obtain transitions ($\bm{s}_t,\bm{a}_t,r_t,\bm{s}_{t+1}$), and collects all transitions into an \textit{experience reply buffer} $R$. Secondly, the agent randomly sample a mini-batch data from $R$ to update two networks.  
\textcolor{black}{Finally, the DCS receives an action that is produced by the actor network $\pi(\bm{s}_t)$ and further fine tuned by the safe layer. The convergence of the DRL algorithm has been proved mathematically \cite{singh2000convergence}.}

Using the experience reply buffer $R$, randomly sampled data keeps weak correlationship with each other, which effectively avoids over-fitting of the two networks. The update rule for actor network $\theta^{\pi}$ is given in its gradient direction as \cite{DDPG2016}:
\begin{align}
	&\bigtriangledown_{\theta^{\pi}}J^{\pi}=\mathbb{E}[\bigtriangledown_{\bm{a}}Q(\bm{s},\pi(\bm{s}))\bigtriangledown_{\theta^{\pi}}\pi(\bm{s})],\label{eqn_drl_7}
\end{align}
where the gradient of $Q$ needs to be estimated through the critic network. Moreover, because the sampled transitions from $R$ are all guided by policy $\pi$, the Monte-Carlo approach is adopted to give an un-biased estimate of Eq. (\ref{eqn_drl_7}) as:
\begin{align}
	&\bigtriangledown_{\theta^{\pi}}J^{\pi}\approx\frac{1}{K}\sum\nolimits_{k=1}^{K}\bigtriangledown_{\bm{a}}Q(\bm{s}_{k},\pi(\bm{s}_{k}))\bigtriangledown_{\theta^{\pi}}\pi(\bm{s}_{k}),\label{eqn_drl_10}
\end{align}
where $k$ is the index of samples; $K$ is the size of the sampled mini-batch data set $\mathcal{K}$. For the critic network $\theta^{Q}$, the mean squared error (MSE) is used as the loss function: 
\begin{align}
	&L=\frac{1}{K}\sum\nolimits_{k=1}^{K}[y_{k} - Q(\bm{s}_{k},\bm{a}_{k})]^{{2}},\label{eqn_drl_8}
\end{align}
where $y_{k}$ is the target value of $Q(\bm{s}_{k},\bm{a}_{k})$ and needs to be estimated. To stabilize the training process and guarantee the convergence, the target $y_{k}$ should not change frequently. According to the Bellman Expectation Equation of Eq. (\ref{eqn_drl_4}), two target networks ($Q',\pi'$), copies of ordinary networks ($Q,\pi$), are designed to calculate $y_{k}$ as:
\begin{align}
	&Q^{\pi}(\bm{s}_{t},\bm{a}_{t})=\mathbb{E}[r_{t}+\gamma Q^{\pi}(\bm{s}_{t+1},\bm{a}_{t+1})],\quad\forall t \in\mathcal{T},\label{eqn_drl_9}\\
	&y_{k}=r_{k}+Q'(\bm{s}_{k+1},\pi'(\bm{s}_{k+1})), \quad \forall k\in \mathcal{K}.\label{eqn_drl_10}
\end{align}
To be more stabilized, the target networks are updated following the running average method, which are given by:
\begin{align}
	&\theta^{Q'}\gets\tau\theta^{Q}+(1-\tau)\theta^{Q'},\label{eqn_ddpg_6}\\
	&\theta^{\pi'}\gets\tau\theta^{\pi}+(1-\tau)\theta^{\pi'},\label{eqn_ddpg_7}
\end{align}
where $\tau$ is the smooth factor, $0\leq\tau\ll1$. Finally, to improve the efficiency of the exploration, an independent noise $\xi_t$ is added to each action subject to the Gaussian distribution $\xi\sim N(0,\sigma^{{2}})$. 
\textcolor{black}{The proposed algorithm is summarized in Table~\ref{ddpg_algorithm}, where the safe layer showed as row 06 will be described in detail in the next subsection.} 

\subsection{Constrained Policy by the Safe Layer}\label{safe_chapter}

\begin{table}
    % \color{red}
	%	\renewcommand{\arraystretch}{1.25}
	\begin{small}
		{
			\caption{Safe-DRL algorithm}
			\label{ddpg_algorithm}
			%		\vspace{-2mm}
			\begin{tabular}{p{0.35cm}p{7.6cm}}
				\hline
				$01$&Initialize the random process $\xi$, the experience reply buffer $R$ and the actor, critic networks $Q(\bm{s},\bm{a}),\pi(\bm{s})$ with weights $\theta^{Q},\theta^{\pi}$, respectively. Initialize corresponding two target networks $Q',\pi'$ with weights $\theta^{Q'}\gets\theta^{Q}$, $\theta^{\pi'}\gets\theta^{\pi}$.\\
				$02$&\textbf{For} episode = $1:1:M$ \textbf{do}\\
				$03$&\begin{adjustwidth}{3mm}{0cm}Receive initial observation state $\bm{s}_{t_0}$.\end{adjustwidth}\\
				$04$&\begin{adjustwidth}{3mm}{0cm}\textbf{For} $t=1:1:T$ \textbf{do}\end{adjustwidth}\\
				$05$&\begin{adjustwidth}{6mm}{0cm}Select DCSs' control action $\bm{a}_{t_0+t}=\pi(\bm{s}_{t_0+t})+\xi_{t_0+t}$.\end{adjustwidth}\\
				$06$&\begin{adjustwidth}{6mm}{0cm} fine tune $\bm{a}_{t_0+t}$ by the safe layer.\end{adjustwidth}\\
				$07$&\begin{adjustwidth}{6mm}{0cm}Execute the action $\bm{a}_{t_0+t}$, then obtain the reward $r_{t_0+t}$ and the next state $\bm{s}_{t_0+t+1}$. \end{adjustwidth}\\
				$08$&\begin{adjustwidth}{6mm}{0cm}Collect the transition ($\bm{s_{t_0+t}},\bm{a_{t_0+t}},\bm{r_{t_0+t}},\bm{s_{t_0+t+1}}$) to $R$, and randomly sample a mini-batch data from $R$. \end{adjustwidth}\\
				$09$&\begin{adjustwidth}{6mm}{0cm}Update the actor and critic networks by (\ref{eqn_drl_10}) and (\ref{eqn_drl_8}). \end{adjustwidth}\\
				$10$&\begin{adjustwidth}{6mm}{0cm}Update the two target networks by (\ref{eqn_ddpg_6})-(\ref{eqn_ddpg_7}). \end{adjustwidth}\\
				$11$&\begin{adjustwidth}{3mm}{0cm} \textbf{Endfor}\end{adjustwidth}\\
				$12$ &\textbf{Endfor}\\
				\hline
		\end{tabular}}
	\end{small}
	\vspace{-4mm}
\end{table}

As shown in step 06 of Table~\ref{ddpg_algorithm}, the action $\bm{a}_t$ needs to be fine tuned by the safe layer before being executed, which aims to guarantee the critical constraint of the operating power in Eq. (\ref{eqn_drl_6}). 
\textcolor{black}{The proposed safe layer achieves the required power cap to assure the high-quality performance in the operating reserve, which effectively makes the DRL agent's control results more reliable in practice.}

In each time step $t\in\mathcal{T}$, according to the output action $\bm{a}_{t}=\Delta \bm{m}^{\text{I}}_{t}$, the next mass flow rate of buildings $\bm{m}^{\text{I}}_{t+1}$ can be obtained as:
\begin{align}
&\bm{m}^{\text{I}}_{t+1}=\bm{m}^{\text{I}}_{t}+\Delta \bm{m}^{\text{I}}_{t},\quad\forall t\in\mathcal{T}. \label{eqn_safe_0}
\end{align}
Thus, based on energy balance Eqs. (\ref{eqn_ch_1})-(\ref{eqn_ch_2}), the power consumption at the next state is calculated as:
\begin{align}
	P^{\text{ch}}_{t+1}&=\sum\nolimits_{i\in \mathcal{I}}{m}^{\text{I}}_{i,t+1}\Theta_{t},\quad\forall t\in\mathcal{T},\label{eqn_safe_2}
\end{align}
where $\Theta_{t}=\frac{1}{\text{COP}}[c^{\text{w}} (T^{\text{ch,r}}_{t}-T^{\text{ch,s}})]$ is the known parameter related with the return water temperature. Then, if the power consumption satisfies the constraint $P^{\text{ch}}_{t+1}\leq P^{\text{cap}}$, the action $\Delta \bm{m}^{\text{I}}_{t}$ will be executed directly; Otherwise, the action $\Delta \bm{m}^{\text{I}}_{t}$ should be optimized to decrease the operating power. 

To address this issue, we propose the following linear mapping rule to adjust $\Delta \bm{m}^{\text{I}}_{t}$ as:
\begin{align}
	&\textcolor{black}{\Delta\tilde{\bm{m}}^{\text{I}}_{t} 
	= \Delta\bm{m}^{\text{I}}_{t} + \mu_t\Delta \bm{m}^{\text{I}}_{t}+\upsilon_t\bm{m}^{\text{I}}_{t},\quad\forall t\in\mathcal{T},\label{eqn_safe_3}
	}
\end{align}
\textcolor{black}{where $\mu_t$ and $\upsilon_t$ are the correction coefficients for adjusting the action $\Delta \bm{m}^{\text{I}}_{t}$, and $\mu_t,\upsilon_t\leq 0$; $\Delta\tilde{\bm{m}}^{\text{I}}_{t}$ is the updated action from the original agent's output and will finally be executed in DCS.} 
When $\mu_t$ and $\upsilon_t$ are close to 0, the last two correction terms in Eq.~(\ref{eqn_safe_3}) will take small function, i.e., the original agent's action $\Delta \bm{m}^{\text{I}}_{t}$ will not be adjusted too much by the safe layer.
By contrast, when $\mu_t$ and $\upsilon_t$ are negative and far from 0, the original agent's action $\Delta \bm{m}^{\text{I}}_{t}$ will be adjusted significantly. 
\textcolor{black}{In other words, the safe layer is not only a simple saturation function, but also needs to train the agent to converge.
If the decision from the agent $\Delta \bm{m}^{\text{I}}_{t}$ is changed quite a lot by the safe layer, which probably decreases the agent's training efficiency and even leads to the failure of its convergence.} Therefore, the coefficients $\mu_t$ and $\upsilon_t$ are expected to be large and close to 0. On this basis, the two coefficients $\mu_t$ and $\upsilon_t$ can be optimized by following linear programming:
\begin{align}
	&\max_{\mu_t,\upsilon_t}~ \mu_t+\upsilon_t,\label{eqn_safe_obj}\\
	\text{s.t.:}~
	&\sum\nolimits_{i\in \mathcal{I}} (\mu_t\Delta {m}^{\text{I}}_{i,t}+\upsilon_t m^{\text{I}}_{i,t}) \Theta_{t}\leq P^{\text{cap}},~~\forall t \in\mathcal{T},\label{eqn_cons_1}\\
	& \underline{m}_i^\text{I}\leq \mu_t\Delta {m}^{\text{I}}_{i,t} + \upsilon_t m^{\text{I}}_{i,t} \leq \overline{m}_i^\text{I} ,\quad\forall i\in\mathcal{I},\forall t\in\mathcal{T},\label{eqn_cons_2}\\
% 	&  a_t\Delta m^{\text{I}}_{i,t}+b_t m^{\text{I}}_{i,t}\geq0,\quad\forall i\in\mathcal{I},t\in\mathcal{T}, \label{eqn_cons_3}\\
	&\mu_t,\upsilon_t, \leq 0,\quad\forall t\in\mathcal{T},\label{eqn_cons_4}
\end{align} 
where the objective in Eq. (\ref{eqn_safe_obj}) represents the minimum changes on the original agent's action $\Delta \bm{m}^{\text{I}}_{t}$. The constraint in Eq. (\ref{eqn_cons_1}) is to satisfy the required power cap from power systems. 
Inequalities (\ref{eqn_cons_2})-(\ref{eqn_cons_4}) define the domain of parameters $\mu_t,\upsilon_t\leq 0$ and $\underline{m}^{\text{I}}_{i}\leq m^{\text{I}}_{i,t+1}\leq \overline{m}^{\text{I}}_{i}$. \textcolor{black}{The calculation process of the safe layer is illustrated in Table \ref{safe_algorithm} to achieve the fine tuning of the ``unsafe" action.}

%Similarly, $\upsilon_t$ is for tuning the current mass flow rate $\bm{m}^{\text{I}}_{t}$ in equal proportion. When $\upsilon_t$ is equal to 1, the second term of Eq. (\ref{eqn_safe_3}) will not influence the agent decision $\Delta \bm{m}^{\text{I}}_{t}$. When $\upsilon_t$ is less than 1, the current mass flow rate $\bm{m}^{\text{I}}_{t}$ will take function to reduce the agent decision $\Delta \bm{m}^{\text{I}}_{t}$, so as to guarantee the power constraint. Therefore, the coefficient $\upsilon_t$ is also expected to be larger and closer to 1. 

%Further, each building's mass flow rate $m^{\text{I}}_{i,t+1}$ is constrained in DCS by the system's physical capacity as follows:
%\begin{equation}
%	m^{\text{I}}_{i,t+1}\leftarrow\left\{
%	\begin{aligned}	
%		&\overline{m}^{\text{I}}      
%		&\text{if} ~ m^\text{I}_{i,t+1}\ge\overline{m}^{\text{I}}_{i}  \\
%		& \underline{m}_i^\text{I}             
%		&\text{if} ~ m^{\text{I}}_{i,t+1}\le\underline{m}^{\text{I}}_{i} \\
%		& m^{\text{I}}_{i,t+1}    
%		&\text{otherwise} ~\quad\quad\\
%	\end{aligned},
%	~~\forall i\in\mathcal{I},t\in\mathcal{T},\right.
%	\label{eqn_safe_1}
%\end{equation}
%where if the next mass flow rate is larger (or smaller) than the upper (or lower) limit, the maximum (or minimum) is the limit. It is noted that, the power cap is still satisfied in the physical system, because $\underline{m}_i^\text{I} \leq m^{\text{I}}_{i,t+1}$ in (\ref{eqn_cons_2}) so that the new mass flow rate in Eq. (\ref{eqn_safe_1}) can only be limited by maximum value to decrease the actual operating power.

\begin{table}
	%	\renewcommand{\arraystretch}{1.25}
% 	\color{red}
	\begin{small}
		{
			\caption{Safe layer method}
			\label{safe_algorithm}
			%		\vspace{-2mm}
			\begin{tabular}{p{0.35cm}p{7.6cm}}
				\hline
				$01$&Obtain the next mass flow rate $\bm{m}^{\text{I}}_{t+1}$ and operating power $P^{\text{ch}}_{t+1}$ by (\ref{eqn_safe_0}),~(\ref{eqn_safe_2}).\\
				$02$&\textbf{If} $P^{\text{ch}}_{t+1}\leq P^{\text{cap}}$ \textbf{then}: execute $\Delta\bm{m}^{\text{I}}_{t}$ directly;\\
				$03$&\textbf{Else}\\
				$04$&\begin{adjustwidth}{3mm}{0cm}Solve the optimal coefficients $\mu_t$ and $\upsilon_t$ by (\ref{eqn_safe_obj})-(\ref{eqn_cons_4}); \end{adjustwidth}\\
				$05$&\begin{adjustwidth}{3mm}{0cm}Optimize the next mass flow rate $\Delta\bm{m}^{\text{I}}_{t}$ using (\ref{eqn_safe_3}); \end{adjustwidth}\\
				$06$&\begin{adjustwidth}{3mm}{0cm}Execute the fine tuned mass flow rate $\Delta\tilde{\bm{m}}^{\text{I}}_{t}$. \end{adjustwidth}\\
				$06$&\textbf{End}\\
				\hline
			\end{tabular}}
	\end{small}
	\vspace{-4mm}
\end{table}
\color{black}

\begin{remark}
    The mass flow rate is fine tuned by a mapping rule in Eq. (\ref{eqn_safe_3}) to satisfy the power constraint, which maintains all the buildings' relative relation. In this way, the influence to the agent's training process is linear and feasible to learn, which guarantees the convergence of the agent's policy iteration.
\end{remark}

\subsection{Self-adaptive Target Method}\label{safe-layer}
\textcolor{black}{After providing operating reserve, DCS stops following power systems' regulation signals and enters the power recovery stage. Thus the power cap constraint in Eq. (\ref{eqn_drl_6}) is relaxed and the DCS tends to recover buildings' comfort temperature as soon as possible.} However, a too rapid recovery of the temperature may cause an instantaneous increase in the power consumption, called ``power rebound". It may lead to a new power peak and cause stability problems for power systems. 
\textcolor{black}{In some extreme cases, the large load current brought by the power rebound could cause the melting of overhead lines, which damages power system security considerably \cite{rebound2016}.}
To avoid the ``unsafe" power rebound, we further propose a self-adaption target method combined to the proposed safe-DRL scheme to achieve a smooth recovery, as follows: 
\begin{align}
	&r_{t}=-\mathbb{E}_{i\sim\mathcal{I}}[|\Delta T_{i,t+1}-\varphi_{i,t+1}|],\quad\forall t\in[t_1,t_2],\label{eqn_rebound_1}
\end{align}
where $r_{t}$ is the reward of the indoor temperature in the recovery stage; $\varphi_{i,t}$ is the self-adaptive factor; $t_1$ is the end time of the power reduction stage and also the beginning time of the power recovery stage; $t_2$ is the required time for recovering the indoor temperature to the set value.

The reward $r_{t}$ in Eq. (\ref{eqn_rebound_1}) is different from the definition during the reduction stage in Eq. (\ref{eqn_drl_3}). Because Eq. (\ref{eqn_rebound_1}) considers not only the buildings' set values, but also the self-adaptive factor $\varphi_{i,t}$
to design a expected temperature-decreasing trend. In this way, the sharp increase of the DCS's operating power can be alleviated. We propose the following configuration method for the self-adaptive factor\footnote{The configuration principle is to make the indoor temperature recover to 50\% of $\Delta T_{i,t_{1}}$ when the time goes halfway, i.e.,  $t=t_1+\frac{1}{2}(t_2-t_1)$.}: 
\begin{align}
&\varphi_{i,t}=\frac{\Delta T_{i,t_{1}}}{1+e^{\lambda[\frac{t-t_{1}}{t_{2}-t_{1}}-\frac{1}{2}]}},\quad\forall i\in\mathcal{I},\forall t\in[t_1,t_2],
\end{align}
where $\lambda$ is determined according to the required recovery extent of the indoor temperature at time $t_2$. For example, when $\lambda$ is set as 6, the recovery extent of the indoor temperature can reach 95\% of $\Delta T_{i,t_{1}}$ at time $t_2$. Therefore, we can set the values of $\lambda$ and $t_2$ to obtain the self-adaptive factor $\varphi_{i,t}$.
\begin{figure}
    \vspace{-6mm}
	\centering
	\includegraphics[width=1\columnwidth]{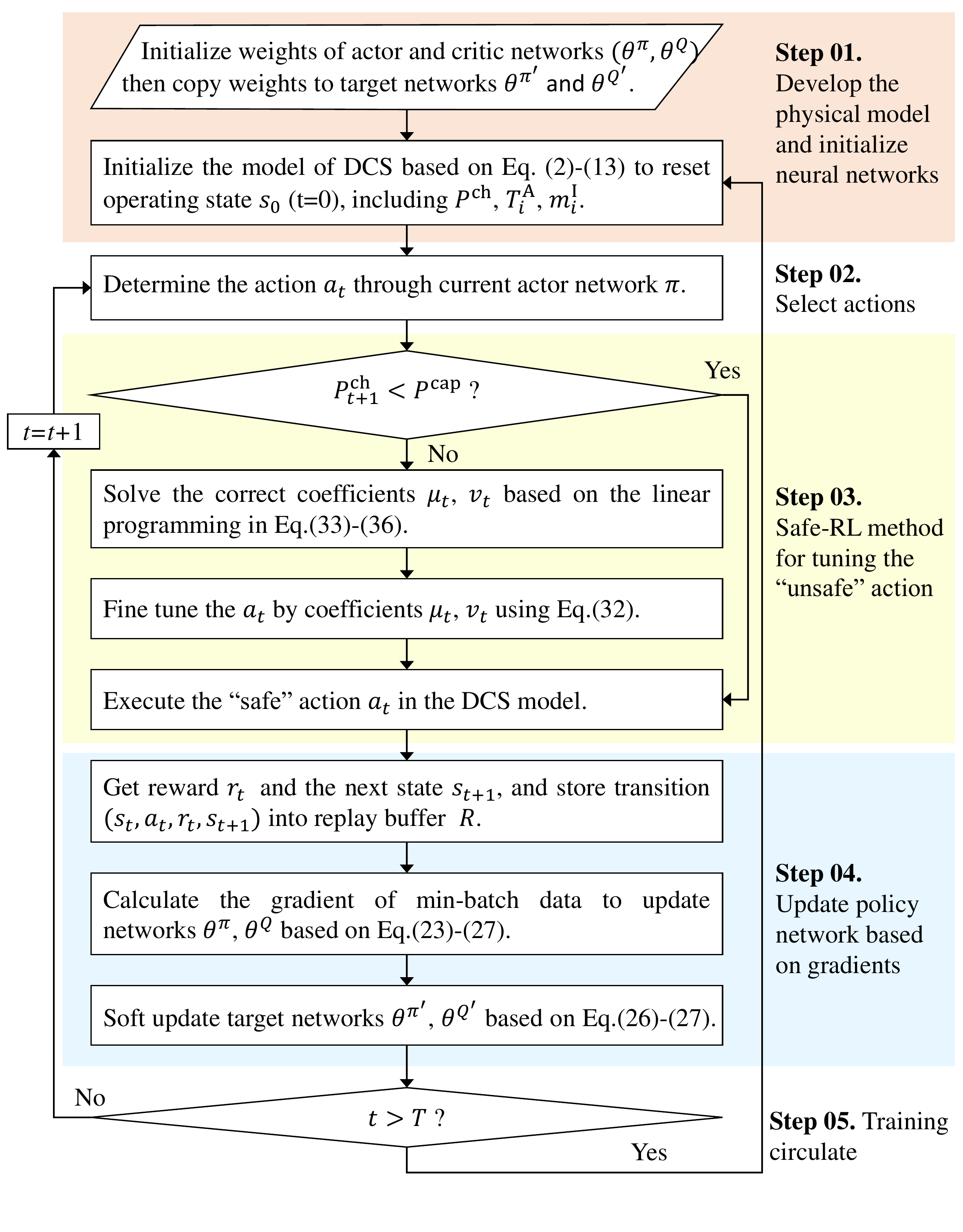}
	\vspace{-7mm}
	\caption{\textcolor{black}{Step-by-step calculation process.}}
	\label{fig_step_by_step}
	\vspace{-4mm} 
\end{figure}

Moreover, in order to constrain the increased operating power
during the recovery stage strictly, we also design a safe layer for the agent, similar with that during the reduction stage in Eq. (27)-(33). The difference is that the $P^\text{cap}$ in Eq. (31) is replaced by the power consumption $P^{\text{ch}}_{t_{0}}$ at time $t_{0}$, given by:
\begin{align}
	&P^{\text{ch}}_{t}\leq\overline{P}^{\text{ch}}=P^{\text{ch}}_{t_{0}},\quad\forall  t \in[t_1,t_2],\label{eqn_rebound_2}
\end{align} 
where $\overline{P}^{\text{ch}}$
is the upper limit of the operating power during the recovery stage. 
\textcolor{black}{The training process of the safe-DRL agent is organized in Fig.~\ref{fig_step_by_step} to show the step-by-step calculation process. It includes 5 steps, where the first step is to develop the environment based on Eqs. (2)-(13) and initialize the policy; the second step is to select an action according to the policy; the third step is to tune the ``unsafe" action by the proposed safe layer as Table II; the fourth step is to update the policy according to samples.}

\begin{remark}
    The proposed self-adaptive target method in Eq. (\ref{eqn_rebound_1}) can regulate the DCS operating power to avoid the power rebound in the power recovery stage and minimize the buildings' comfort impacts.
\end{remark}

%\begin{figure}
%	\centering
%	\includegraphics[width=1\columnwidth]{fig_rebound_process.pdf}
%	\vspace{-23mm}
%	\caption{The concepts of two control stages}
%	\label{rebound_process}
%\end{figure}

%\begin{figure}[h]
%	\setlength{\abovecaptionskip}{0.cm}
%	\setlength{\belowcaptionskip}{-0.cm}
%	\centering
%	\includegraphics[width=1\columnwidth]{fig_temp_dec.pdf}
%	\vspace{-29mm}
%	\caption{Curves for the function $\varphi^{i}$}
%	\label{temp_dec}
%\end{figure}

\section{Case Studies}\label{dcs_case}
\subsection{Test System}
\textcolor{black}{The test system is modelled based on a realistic DCS in Hengqin, China, following its technical guidelines (the 4th Edition) \cite{guide2016}.}
The total installed cooling capacity in the energy station is 41,000 RT ($\approx$144 MW) with COP=5.5. The designed supply and return water temperature in two loops \textcolor{black}{at time $t$=0} is $T^{\text{ch,s}}=\text{3}$ $^\circ$C,
\textcolor{black}{$T^{\text{I,r}}_{i,0}=\text{12}$ $^\circ$C, $T^{\text{II,s}}_{i,0}=\text{13}$ $^\circ$C, $T^{\text{II,r}}_{i,0}=\text{18}$ $^\circ$C,} respectively. In addition, based on the national standard in China \textcolor{black}{(JGJ 134-2010, GB 12021.3-2010, GB 31349-2014)}, the following parameters are designed as \textcolor{black}{$k^{\text{HE}}_i=\text{4.5}$ kW/(m$^{{2}}\cdot^\circ$C),} 
$U^{\text{O-A}}=\text{0.0036}$ kW/(m$^{{2}}\cdot^\circ$C), $c^{\text{w}}=\text{4.2}$ kJ/(kg$\cdot^\circ$C), $c^{\text{A}}=\text{1.005}$ kJ/(kg$\cdot^\circ$C) and $\rho^{\text{A}}=\text{1.205}$ kg/m$^{3}$. The efficiency coefficients of heat exchanging process between different loops are set as $\eta^{\text{I}}_i=\text{0.9}$, $\eta^{\text{II}}_i=\text{0.9}$, respectively. 
\textcolor{black}{The heat transfer coefficient of supply water $\eta^\text{pipe}$ is 0.95.}
The air mixing proportion is set as $\alpha_i=\text{0.1}$.

\begin{figure}%[H]
	 %\vspace{-4mm}
	\centering
	\includegraphics[width=1\columnwidth]{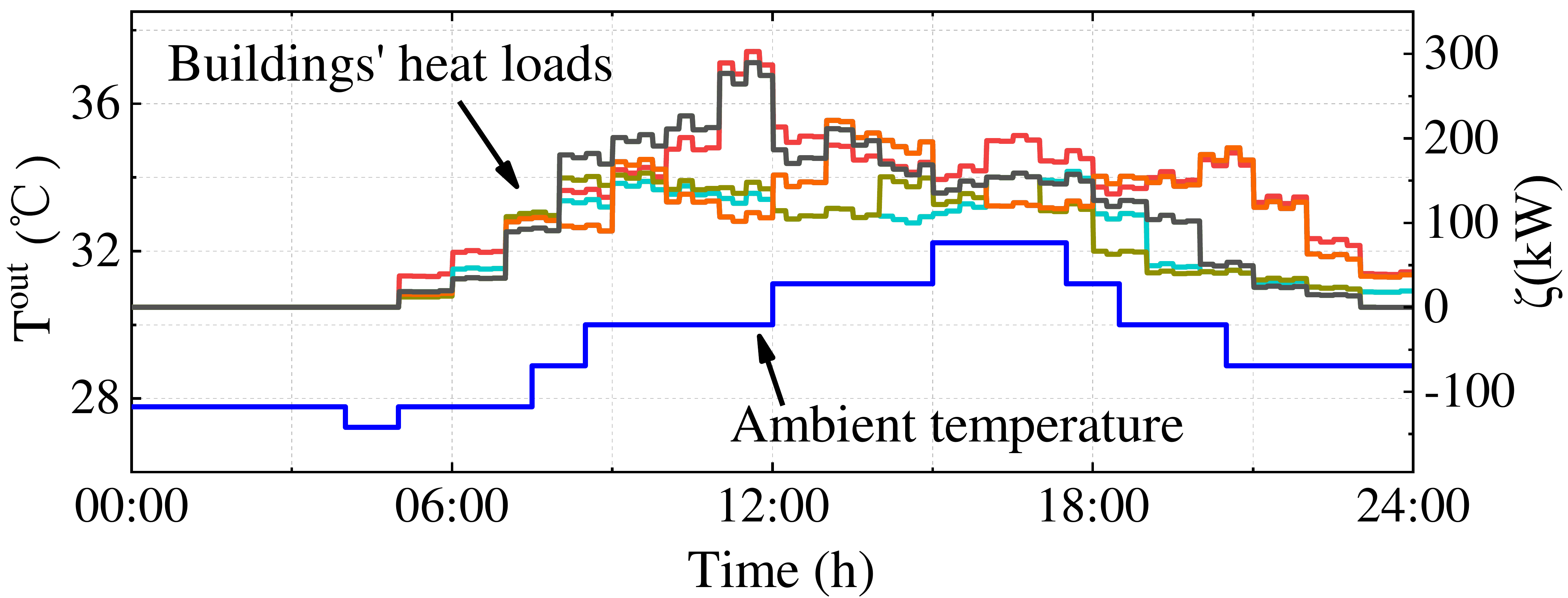}
	\vspace{-5mm}
	\caption{\textcolor{black}{The ambient temperature and buildings' heat loads.}}
	\label{fig_Tout_ee}
	\vspace{-4mm}
\end{figure}

The DCS in Hengqin provides cooling services for 12 buildings. The maximum value of the mass flow rate $\overline{m}^{\text{I}}_{i}$ ranges from 600 kg/s to 1,200 kg/s in different buildings, and the corresponding minimum value $\underline{m}^{\text{I}}_{i}$ is $\text{3}\%$ of $\overline{m}^{\text{I}}_{i}$. Each building's floor area $A_i^\text{S}$ and its set temperature $T^{\text{set}}_{i,t}$ are distributed in 100,000$\sim$300,000 m$^{2}$, and 20$\sim$23 $^\circ$C, respectively. The maximum deviation of the required comfortable indoor temperature is $\pm$1 $^\circ$C. 
\textcolor{black}{Moreover, the ambient temperature $T^{\text{out}}_{t}$ and each building's heat load $\zeta_{i,t}$ adopt the realistic data in Hengqin, from June 1, 2020 to August 31, 2020 (one typical day's profiles are shown in Fig.~\ref{fig_Tout_ee}).} 

\begin{figure}%[H]
	 %\vspace{-4mm}
	\centering
    \includegraphics[width=1\columnwidth]{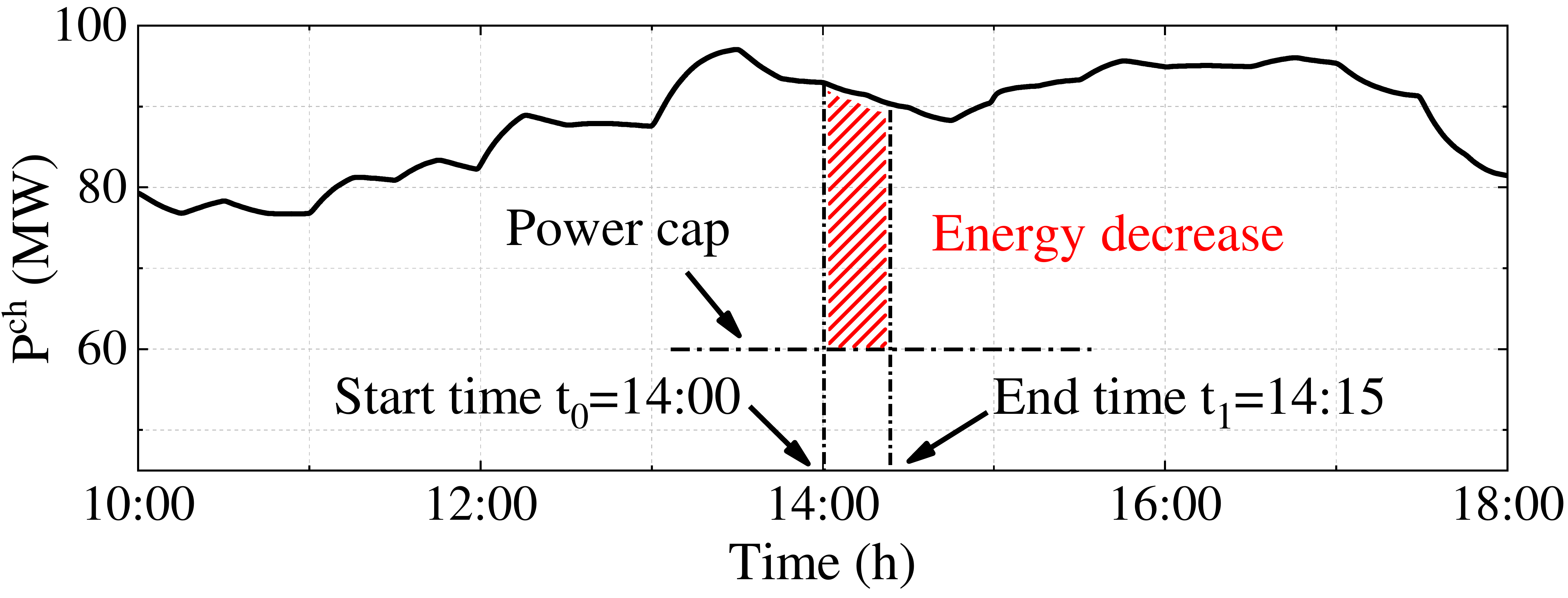}
	\vspace{-5mm}
	\caption{\textcolor{black}{The original power consumption of DCS.}}
	\label{fig_case_daily}
	\vspace{-4mm}
\end{figure}

The control objective of DCS is to provide operating reserve from 14:00pm to 14:15pm, as shown in Fig.~\ref{fig_case_daily}.\footnote{\textcolor{black}{Note that these experimental settings are for illustrative purpose. In practice, the service duration, the operating reserve period and the power cap $P^{\text{cap}}$ are determined by the system operator. The effectiveness of the proposed methodology is not affected by these parameter settings.}}
The black curve is the original power consumption, and regarded as the power baseline before regulation. The red shadow area is the required decrease of energy consumption, and the operating power should be lower than the power cap $P^{\text{cap}}$=60 MW during this period. In the recovery stage, the new power cap is set as the peak power of the baseline, i.e., $\overline{P}^{\text{ch}}$ = 96 MW. 

\subsection{Benchmarks}

\textcolor{black}{To validate the superiority of the proposed safe-DRL scheme, we implement another two centralized control methods as our benchmarks: the traditional proportional-integral (PI) controller \cite{PI_controller} and the DRL controller \cite{mnih2015human}.}
\textcolor{black}{Here, the superiority includes providing a higher-quality service and preventing the power rebound with the minor impacts on buildings' temperature comforts.}

In the PI method, the control signal is determined by the feedback of both the power cap violation and indoor temperature comfort\footnote{\textcolor{black}{A more complex version of PI controller is PID controller, in which a D parameter is added to reflect the differential control process. We have also tested PID controller's performance for DCS control. However, our experiments show that PID does not obviously outperform PI (but with a more complex structure). Furthermore, when adopting a PID controller with a large D parameter, there may exist a strong noise in the controlling results. Hence, we only include PI controller in this paper to save space.}}.
The buildings' mass flow regulation at each time step $t$ can be expressed as:
\begin{align}
    \Delta m^\text{ch}_{t} = &\text{P}^\text{ch}(P^\text{ch}_t-P^\text{ch}_{t-1}) +  \text{I}^\text{ch}(P^\text{ch}_t-P^\text{cap}),\\
    \Delta m^\text{I}_{i,t} =
    &\left [ 
    \text{P}_i(T^\text{A}_{i,t}-T^\text{A}_{i,t-1}) + \text{I}_i(T^\text{A}_{i,t}-T^\text{set}_{i,t}) 
    \right ]\notag\\
    &+m^\text{I}_{i,t-1}\Delta m^\text{ch}_{t}/\sum\nolimits_{i\in\mathcal{I}}m^\text{I}_{i,t-1},
    \label{PI_control_1}
\end{align}
where $\text{P}^\text{ch}$, $\text{I}^\text{ch}$ are parameters of the PI controller in pipelines to follow power caps; $\text{P}_i$, $\text{I}_i$ are the parameters of PI controllers in buildings to follow set indoor temperatures. Eq.~(\ref{PI_control_1}) means the regulation of the total mass flow is achieved by adjusting each building proportionally. During the power reduction stage, parameters are set as $\text{P}^\text{ch}$=0.2, $\text{I}^\text{ch}$=0.02. During the power recovery stage, the power constraint is relaxed and $\text{P}^\text{ch}$=$\text{I}^\text{ch}$=0\footnote{\textcolor{black}{The tuning rule is to satisfy the daily operating and maintain buildings' comfortable set temperatures.}}. 

\textcolor{black}{In the traditional DRL method, it does not have the safe layer so that the power constraint $P^\text{cap}$ is considered as a penalty item in its reward function, which is formulated as:
\begin{align}
	r_{t+1}=
	&- \theta^\text{r}\mathbb{E}_{i\in\mathcal{I}}[|\Delta T_{i,t+1}|] 
	- \sigma^{2}_{i\in\mathcal{I}}[\Delta T_{i,t+1}]\notag\\
	&- \theta^\text{p}|P^\text{ch}_{t+1}-P^\text{cap}|
	,\quad\forall t\in\mathcal{T},\label{eqn_drl_100}
\end{align}
where $\theta^\text{p}$ is the weight factor of the penalty item.}

\subsection{Training Process of the Safe-DRL Agent}
%要加对比图
%要加效率对比
\textcolor{black}{The parameters of the proposed safe-DRL are designed as Table~\ref{DDPG_setting}. 
The key hyper-parameters are designed based on the experience concluded in the  existing literature \cite{mousavi2016deep}, including the discount factor, learning rates, replay buffer capacity, etc..}
The actor and critic networks are composed by one input layer, two hidden layers and one output layer, respectively\footnote{Layer number adopts existing literature models that performs well \cite{thomas2017two}.}.
The neurons number in each hidden layer is set as 128. The Rectified Linear Unit is used as the activation function.

The parameters in DRL (benchmark) adopt the same experimental settings with safe-DRL. The simulation is implemented by Windows system, using PyTorch in Python with an Intel core i7 CPU @3.0 GHz and 16GB memory.
\begin{table}%[h]
	\renewcommand\arraystretch{1} 
	\small %表头的字变小
	\centering
%	\vspace{4mm}
	\caption{Parameters for safe-DRL and DRL methods.}
    \setlength{\tabcolsep}{4mm}%7可随机设置，调整到适合自己的大小为止
	%	\vspace{-2mm}
	\begin{tabular}{ccc}
		\hline
		Symbols&Definitions &Values\\
		\hline
		$\tau$&Target smooth factor&0.005 \\
		$\gamma$&Discount factor & 0.9  \\	
		$|R|$&Replay buffer capacity & 10000  \\
		$\xi$&Exploration noise & 0.3 \\
		$M$&Max episodes & 2500  \\
		$T$&Max step &15\\
		\textcolor{black}{$K$}& \textcolor{black}{Mini batch size} & \textcolor{black}{200}  \\
		\textcolor{black}{$\delta _{\theta^{Q}}$}& \textcolor{black}{Learning rate of critic network $Q$} & \textcolor{black}{0.001}  \\
		\textcolor{black}{$\delta _{\theta^{\pi}}$}&
		\textcolor{black}{Learning rate of actor network $\pi$} & \textcolor{black}{0.0001}  \\
		\textcolor{black}{$\theta^\text{r}$}&
		\textcolor{black}{Weight factor of temp deviations} & \textcolor{black}{0.01}  \\
		\textcolor{black}{$\theta^\text{p}$}&
		\textcolor{black}{Weight factor of power violations} & \textcolor{black}{0.05}  \\
		\hline
	\end{tabular}\label{DDPG_setting}
	%\vspace{-4mm}
\end{table}

\begin{figure}
	\subfigbottomskip=-4pt
	\subfigcapskip=-4pt
	%\vspace{-4mm}
	\centering
	\subfigure[]{\includegraphics[width=0.9\columnwidth]{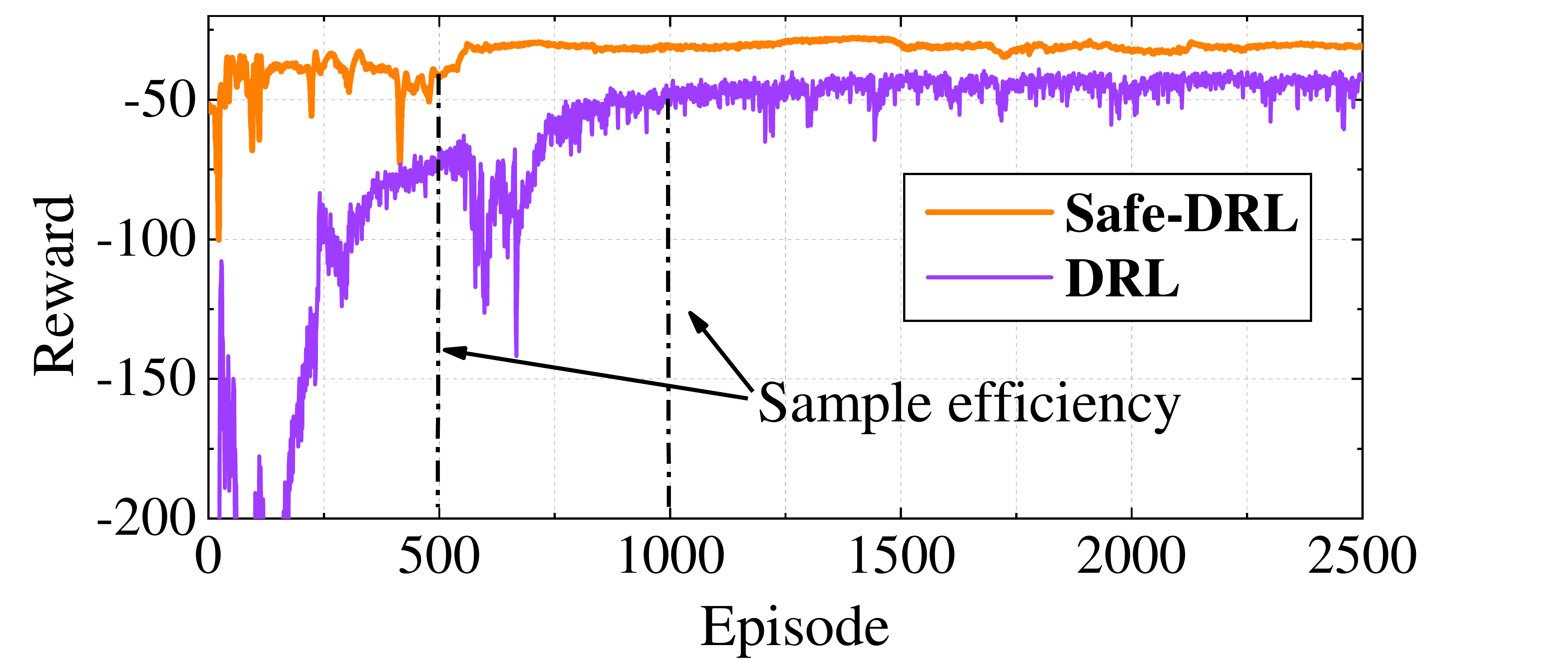}\label{fig_case_training_a}}
	\subfigure[]{\includegraphics[width=0.9\columnwidth]{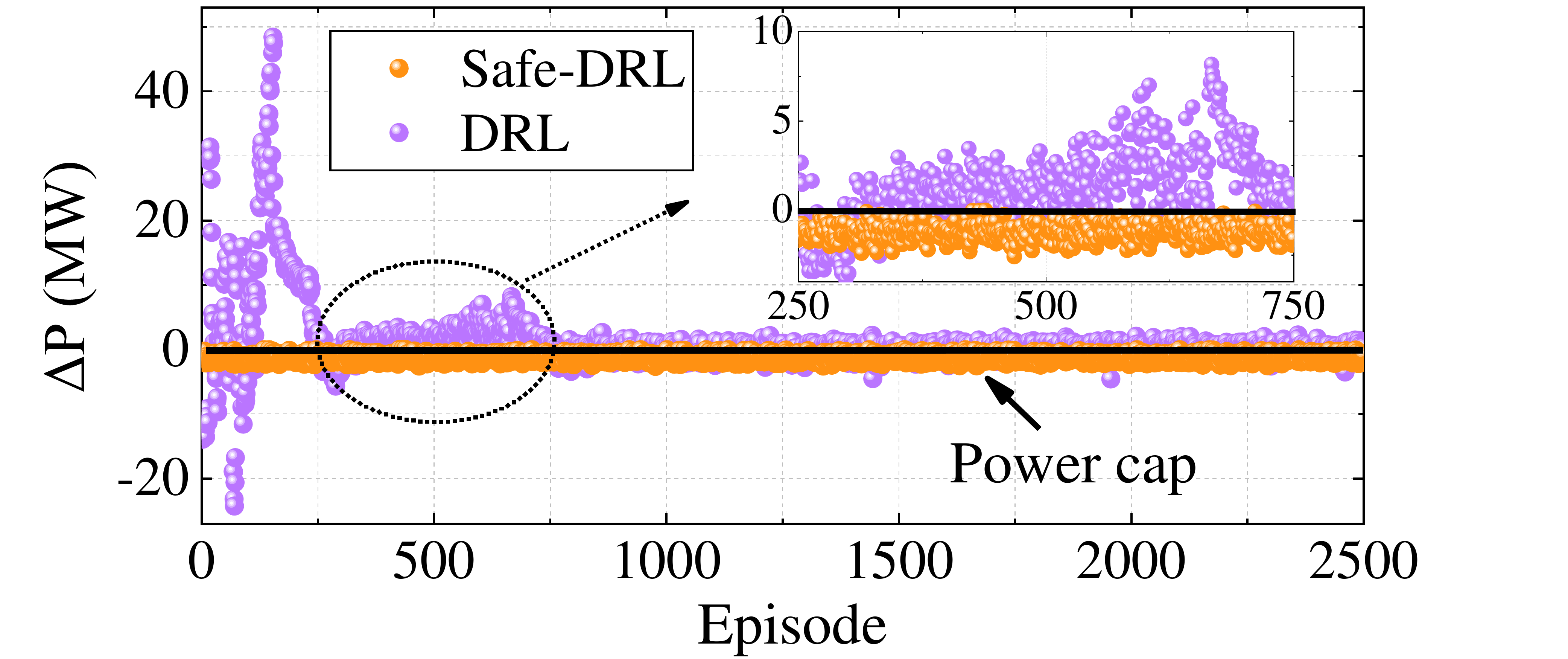}\label{fig_case_training_b}}
	\caption{\textcolor{black}{The training process of the safe-DRL agent. (a) The reward value; (b) The constraint value of operating power.}}
	\label{fig_case_training}
% 	\vspace{-5mm}
\end{figure}

The training process is shown in Fig. \ref{fig_case_training}, and the number of training episode is 2500. Fig.~\ref{fig_case_training_a} presents the reward value for appraising the agent's decision in each episode. 
\textcolor{black}{It can be seen that the rewards in safe-DRL and DRL have oscillations at first because of the unknown knowledge about the training environment (i.e., the DCS). With the increase of training episodes, the rewards converge to their respective stable values, called convergence reward. Then, both of the two agents obtain the optimal policy in Eq.~(\ref{eqn_drl_5}). }

\textcolor{black}{The comparison of training efficiency between the proposed safe-DRL and the traditional DRL method is shown in Table \ref{benchmark_efficiency}, where the efficiency indicator includes sample efficiency, convergence time and convergence reward. Sample efficiency is the estimated minimum number of samples to converge as illustrated in Fig.~\ref{fig_case_training_a}. It can be seen that the proposed safe-DRL needs less sampled 
episodes to converge, so it has higher sample efficiency and shorter convergence time (2.9 mins). Besides, the convergence reward of safe-DRL is larger (-30) than that of DRL (-45), which means the safe-DRL agent can achieve the temperature objective better than DRL.}

\textcolor{black}{Fig.~\ref{fig_case_training_b} shows the constraint violation during the agents' training processes, where $\Delta P=P^\text{ch}-P^\text{cap}$ is the power gap to the required power cap. It can be seen that the power constraint violation is conspicuous and even reaches to over 40MW in DRL, which may harm the stable operation of the power system. However, the operating power can satisfy the power cap strictly in safe-DRL, which proves the effectiveness of the proposed safe-layer.}
Besides, the operating power is quite close to the power cap, because the agent wants to make full use of the allowable power to decrease the indoor temperature deviations. Thus the well-trained agent can be applied to the online control of DCS for providing operating reserve.

\begin{table}
	\small %表头的字变小
	\centering
	\renewcommand\arraystretch{1.2}%设置表格行高为2
%	\vspace{-3mm}
	\caption{\textcolor{black}{The training efficiency results for 2500 episodes.}}
	\setlength{\tabcolsep}{0.5mm}%7可随机设置，调整到适合自己的大小为止
	%	\vspace{-2mm}
	\begin{tabular}{cccc}
		\hline
		Methods &  Sample efficiency &  Convergence time & Convergence reward\\
		\hline
		Safe-DRL & $\sim$500 & 2.9 mins & -30 \\
		DRL & $\sim$1000 & 4.3 mins & -45 \\
		\hline
	\end{tabular}\label{benchmark_efficiency}
	\vspace{-4mm}
\end{table} 

\subsection{Online Control of DCS for Providing Operating Reserve}
\begin{figure}
    \vspace{-5mm}
	\subfigbottomskip=-4pt
	\subfigcapskip=-4pt
	\centering
	\subfigure[]{\includegraphics[width=1\columnwidth]{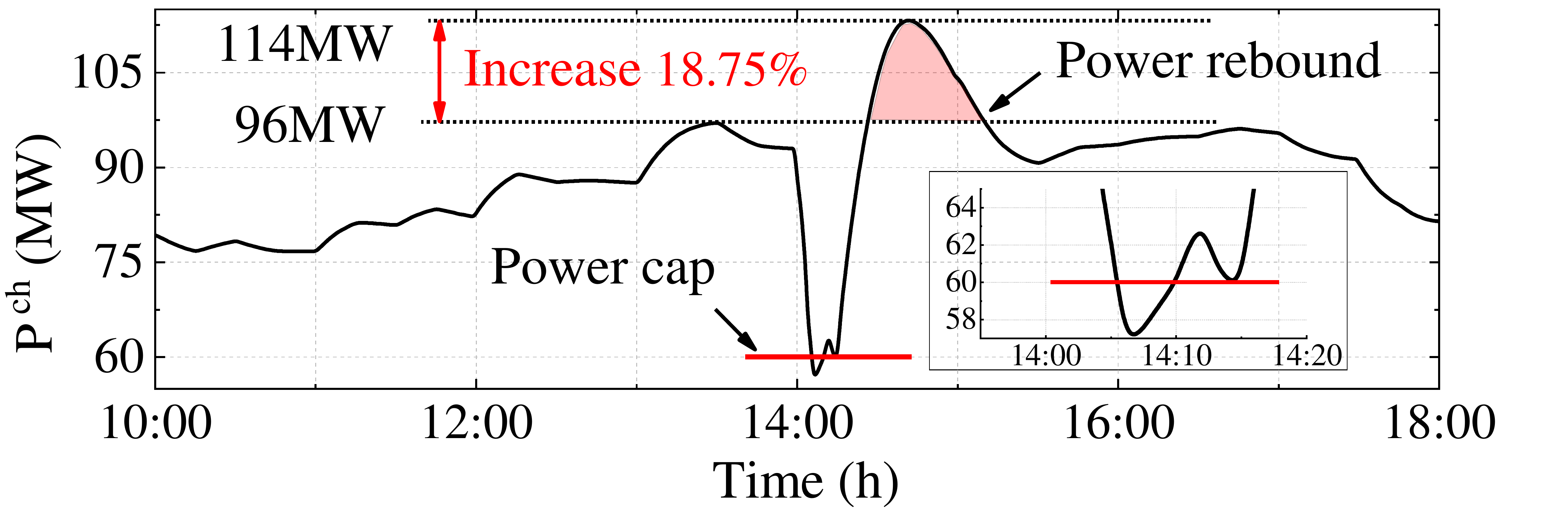}\label{fig_case_power_a}}
	\subfigure[]{\includegraphics[width=1\columnwidth]{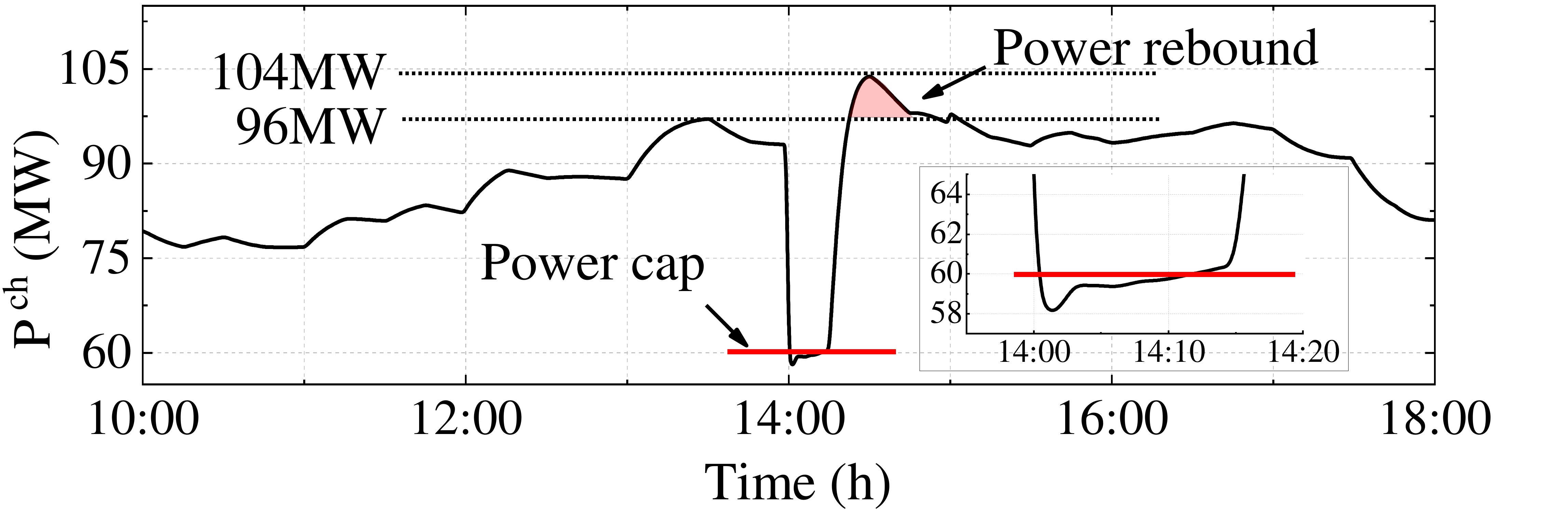}\label{fig_case_power_b}}
	\subfigure[]{\includegraphics[width=1\columnwidth]{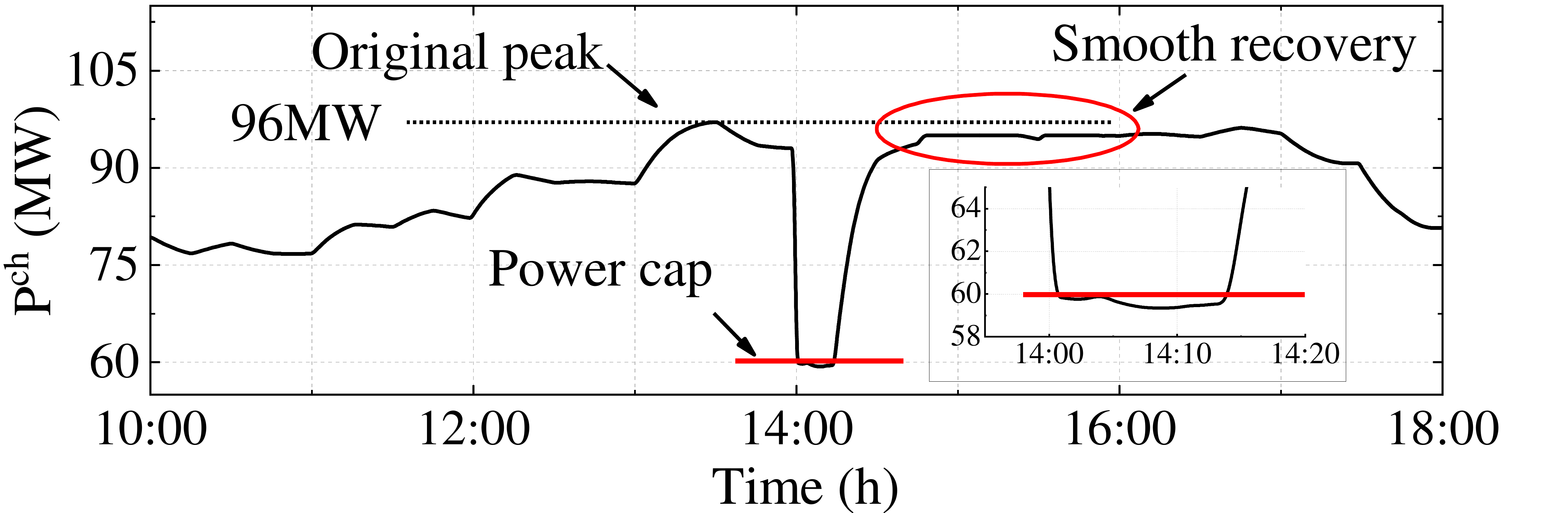}\label{fig_case_power_c}}
	\vspace{-1mm}
	\caption{\textcolor{black}{The control power results of DCS based on (a) PI controller; (b) Traditional DRL method; (c) Safe-DRL method.}}
	\label{fig_case_power}
\end{figure}

For a random case, it is assumed that the power system has the regulation demand at 14:00pm, and sends the regulation signal to the agent to cut down the DCS's operating power to be lower than 60 MW in this dispatch period (15 mins). 
\textcolor{black}{The control power results of DCS for providing operating reserve is shown in Fig.~\ref{fig_case_power}, which applies three different controllers (i.e., PI, the traditional DRL and the proposed safe-DRL) to the system in Fig.~\ref{fig_case_power_a}, Fig.~\ref{fig_case_power_b} and Fig.~\ref{fig_case_power_c}, respectively.}

It can be seen from Fig.~\ref{fig_case_power_a} that DCS operating power is cut down and satisfies the required power after 5 mins.
\textcolor{black}{Because the PI controller is designed based on the feedback, it cannot respond to the changing environment immediately and results in some time delay.}
In Fig.~\ref{fig_case_power_b} and Fig.~\ref{fig_case_power_c}, the DRL and safe-DRL controllers can decrease the operating power more quickly compared with PI controller, where the power reduction is achieved only within 1 min.
Moreover, during the whole dispatch period, the operating power in Fig.~\ref{fig_case_power_a} cannot be maintained below 60 MW and exceeds the required power cap at 14:11 due to the dynamic cooling demand in buildings \textcolor{black}{(e.g., variational heat loads caused by people flows)}. By contrast, the operating power can be controlled under the power cap during all the dispatch period in Fig.~\ref{fig_case_power_c}, which validates the effectiveness of the proposed safe-DRL agent to satisfy power system's critical constraint strictly. 
In Fig.~\ref{fig_case_power_b}, the traditional DRL method can also achieve the required power cap after training, however its training process can not satisfy the constraint.

After the power reduction stage, three controllers in Fig.~\ref{fig_case_power} increases DCS's operating power to restore buildings' indoor temperatures. \textcolor{black}{However, a new peak power 114 MW and 104 MW appears in the recovery stage in Fig.~\ref{fig_case_power_a} and Fig.~\ref{fig_case_power_b}, respectively. They are even much higher than the original daily maximum operating power (96 MW).} This phenomenon may cause a secondary impact on the power system that has just returned to the stable state. By utilizing the proposed safe-DRL method in Fig.~\ref{fig_case_power_c}, the safe-layer limits the peak value during the recovery stage and guarantees the smooth recovery of the operating power without a new peak power rebound.

\begin{table}
	\small %表头的字变小
	\centering
	\renewcommand\arraystretch{1.2}%设置表格行高为2
	\caption{The statistical indicator of the temperature influence to buildings}
	\setlength{\tabcolsep}{0.6mm}%7可随机设置，调整到适合自己的大小为止
	%	\vspace{-2mm}
	\begin{tabular}{cccc}
		\hline
		Methods&Max deviation & Uncomfortable number& Average deviation \\
		\hline
		PI & 1.57 $^\circ$C & 6 & 0.75 $^\circ$C \\
		DRL& 1.18 $^\circ$C  & 2 & 0.80 $^\circ$C \\
		Safe-DRL& 0.93 $^\circ$C  & 0  & 0.85 $^\circ$C \\
		\hline
	\end{tabular}
	\label{temp_campare}
	\vspace{-4mm}
\end{table}

\begin{figure}%[H]
	\vspace{-5mm}
	\subfigbottomskip=-4pt
	\subfigcapskip=-4pt
	\centering
	\subfigure[]{\includegraphics[width=1\columnwidth]{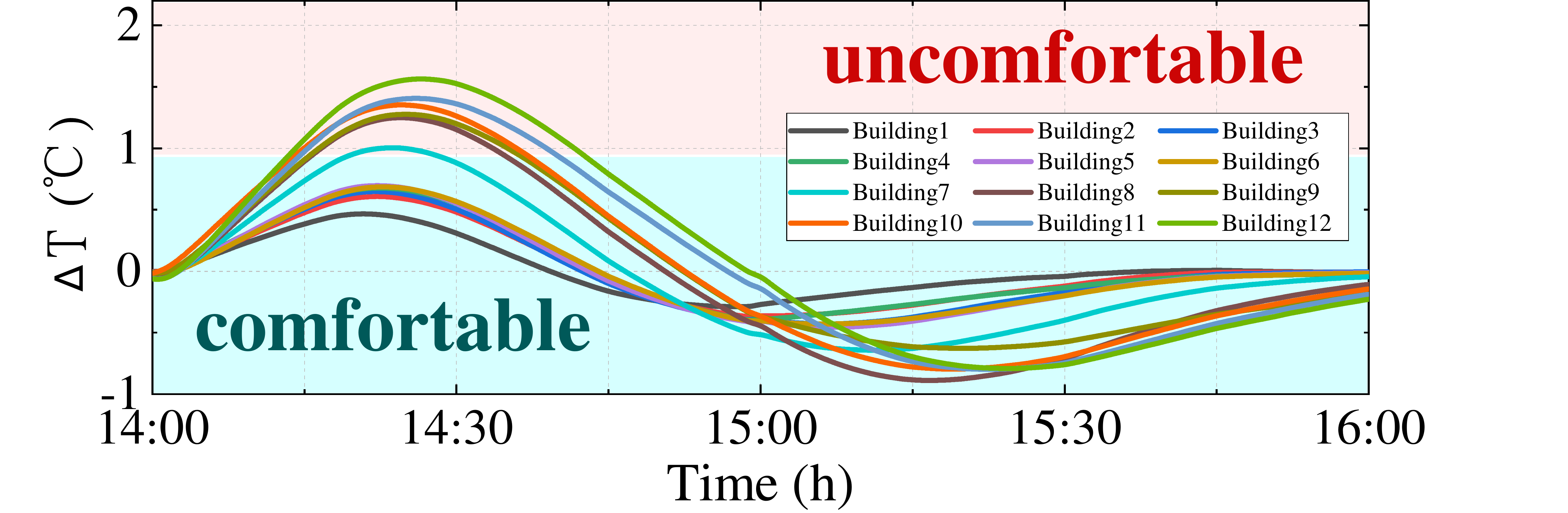}\label{case_temp_delta_a}}
	\subfigure[]{\includegraphics[width=1\columnwidth]{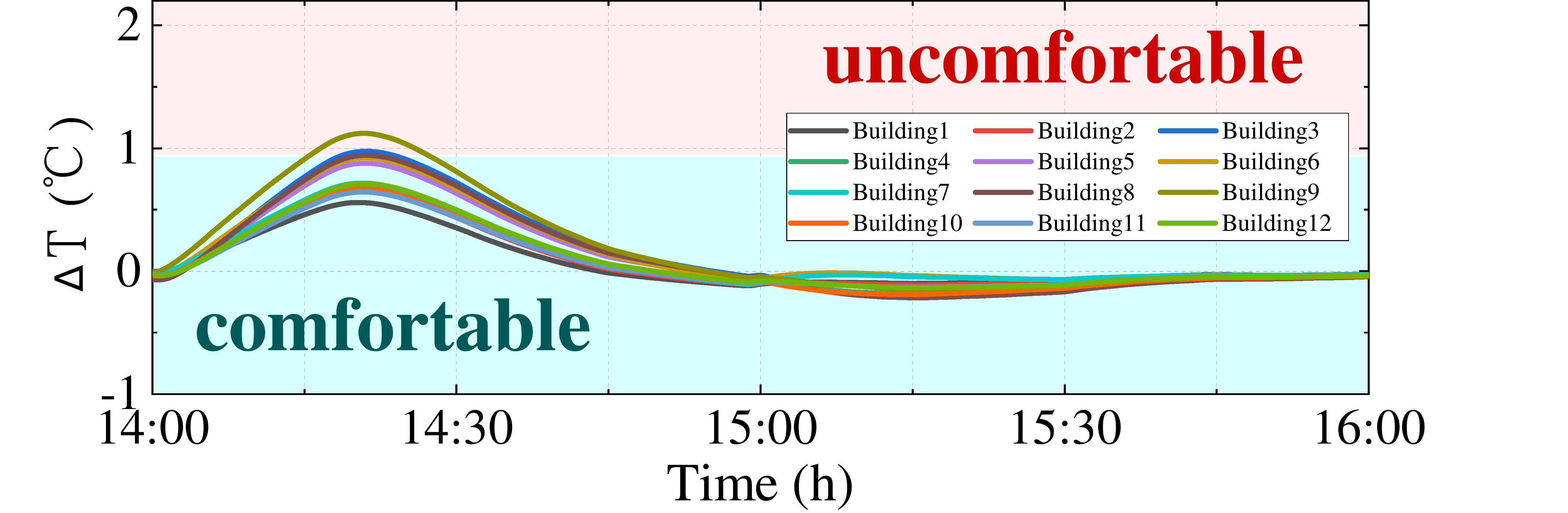}\label{case_temp_delta_b}}
	\subfigure[]{\includegraphics[width=1\columnwidth]{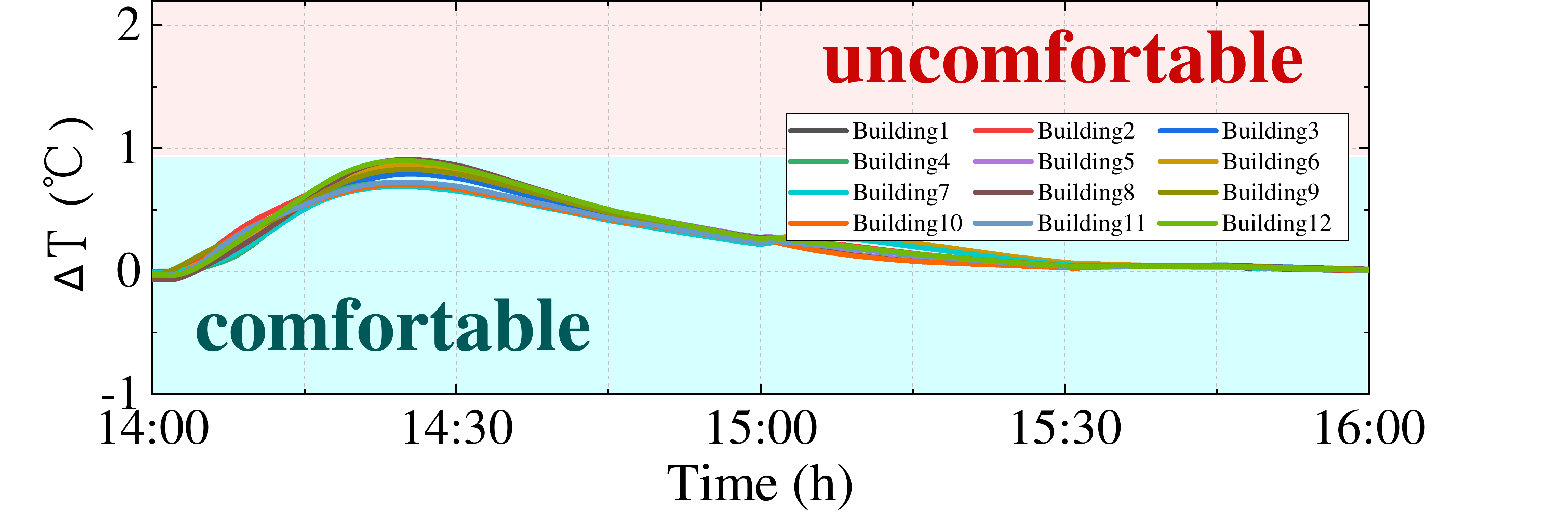}\label{case_temp_delta_c}}
	\caption{\textcolor{black}{The temperature deviation results based on (a) PI controller; (b) Traditional DRL method; (c) Safe-DRL method.}}
	\label{case_temp_delta}
	\vspace{-4mm}
\end{figure}

Moreover, when DCS is controlled to provide operating reserve, building's indoor temperature will get influenced and deviates from its set value, as shown in Fig.~\ref{case_temp_delta}.  
\textcolor{black}{The blue area shows the comfortable temperature range in buildings, and $\Delta T$ denotes each building's temperature deviation. In the power reduction stage, all the buildings' indoor temperatures increase due to the reduction of cooling power supplies.}
In Fig.~\ref{case_temp_delta_a}, more than half of the buildings' indoor temperatures deviate larger than 1 $^\circ$C and enter the uncomfortable area. It means that some buildings get seriously impacted during the regulation process while some others do not.
In Fig.~\ref{case_temp_delta_b}, the DRL method can maintain the temperature comfort better than PI, while some buildings' indoor temperature still exceed the comfortable range.
By contrast, in Fig.~\ref{case_temp_delta_c}, the temperature deviations in different buildings are close and maintained within 1 $^\circ$C by using the safe-DRL controller. 
\textcolor{black}{Although all buildings have different floor areas, heights and heat loads, the temperature influence to each building is always similar. Thus, the safe-DRL method is insensitive to buildings' different models.}
As shown in  Table \ref{temp_campare}, it can be seen the max deviation of buildings in safe-DRL is the smallest, which makes sure more buildings comfortable. However, the average temperature deviation of all the buildings is a little higher (around 0.05\textcelsius) than the other two methods, which is small and can be neglected.
This validates the advantage of the proposed method to regulate each building's mass flow rate dynamically and guarantee their temperature requirements.

\textcolor{black}{It can be seen from Figs.~\ref{fig_case_power}-\ref{case_temp_delta} that the power rebound is more obvious with the PI method, because the indoor temperatures can recover to their set values quickly after the power reduction period and cause the temperature overshoot. Compared with the PI method, the proposed safe-DRL method can control the indoor temperature to recover the set values smoothly. That is to say, the safe-DRL method addresses the power rebound at the cost of a longer recovery time.}

\subsection{Physical Operating States of DCS During Control Process}

\begin{figure}%[H]
	%\vspace{-5mm}
	\subfigbottomskip=-4pt
	\subfigcapskip=-4pt
	\centering
	\subfigure[]{\includegraphics[width=0.88\columnwidth]{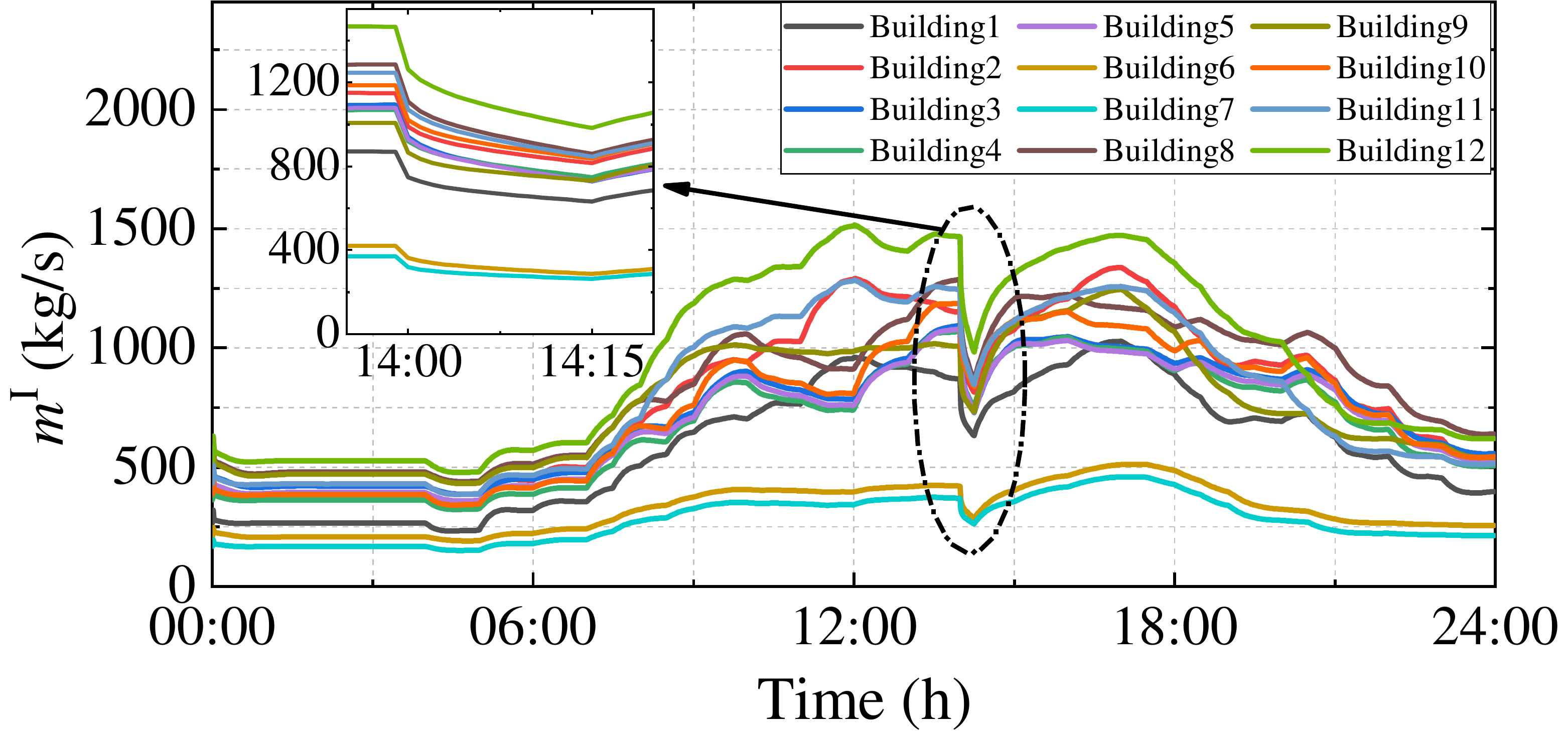}\label{case_physical_water_a}}
	\subfigure[]{\includegraphics[width=0.88\columnwidth]{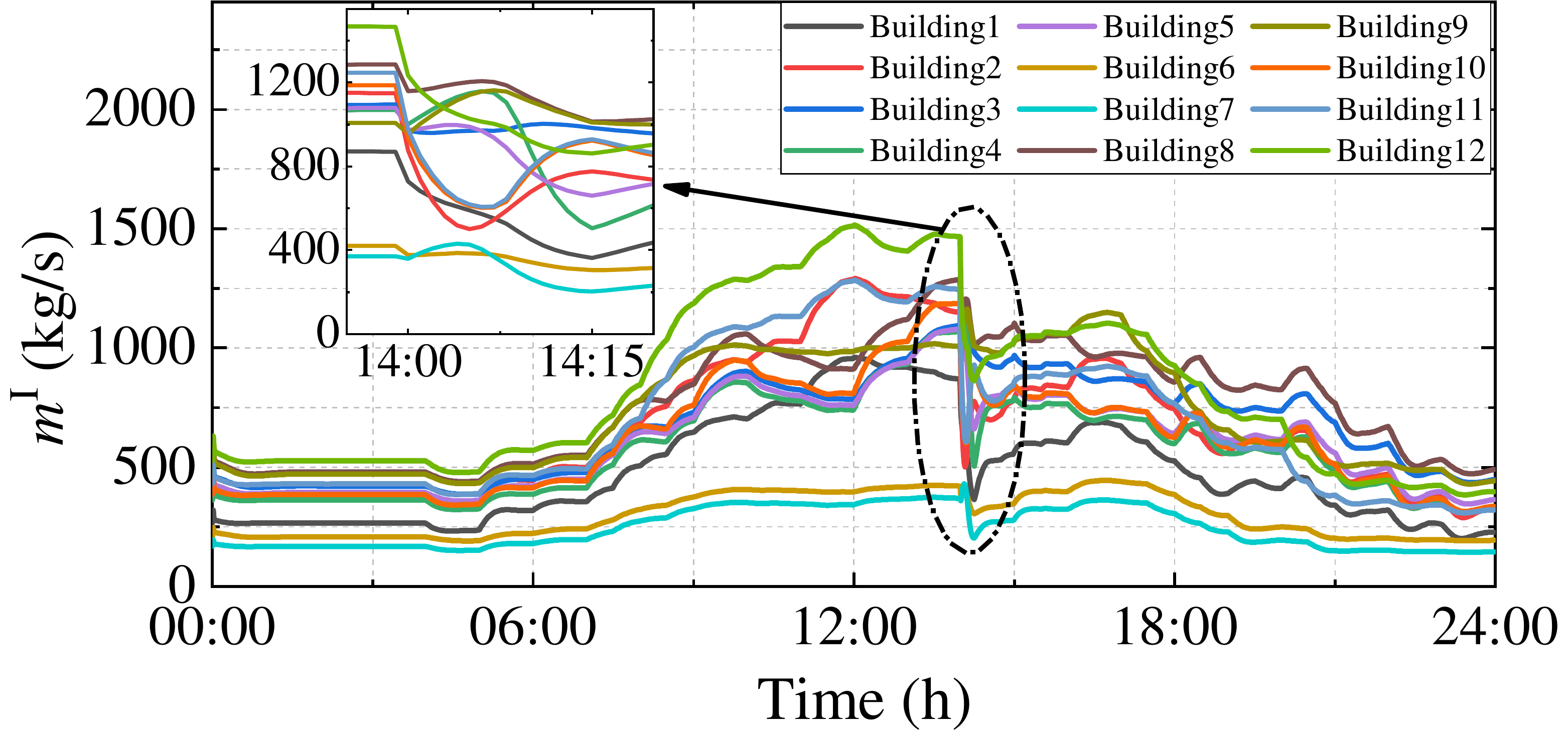}\label{case_physical_water_b}}
	\caption{\textcolor{black}{The water mass flow in buildings based on (a) PI controller; (b) Safe-DRL method.}}
	\label{case_physical_water}
% 	\vspace{-4mm}
\end{figure}

\textcolor{black}{Fig.~\ref{case_physical_water} shows the control results of all the buildings' water mass flows $\bm{m}^\text{I}_t$ in the first water loop. Fig.~\ref{case_physical_water_a} and Fig.~\ref{case_physical_water_b} are obtained based on the PI controller and safe-DRL method, respectively. 
In the power reduction stage, all the buildings' mass flows are decreased slowly based on the feedback with the same proportion in Fig.~\ref{case_physical_water_a}, which is regardless of the differences among buildings. However, mass flows are decreased quickly at first, and then adjusted in different directions for satisfying different cooling demands in heterogeneous buildings in Fig.~\ref{case_physical_water_b}. Therefore, the proposed safe-DRL method can consider different buildings' thermal inertia characteristics and guarantee their comforts.
In the power recovery stage, water mass flows in Fig.~\ref{case_physical_water_a} recover with a faster speed than that in Fig.~\ref{case_physical_water_b}, which makes it easier for the PI controller to cause power rebound.}

\begin{figure}
	%\vspace{-5mm}
	\subfigbottomskip=-4pt
	\subfigcapskip=-4pt
	\centering
	\subfigure[]{\includegraphics[width=0.88\columnwidth]{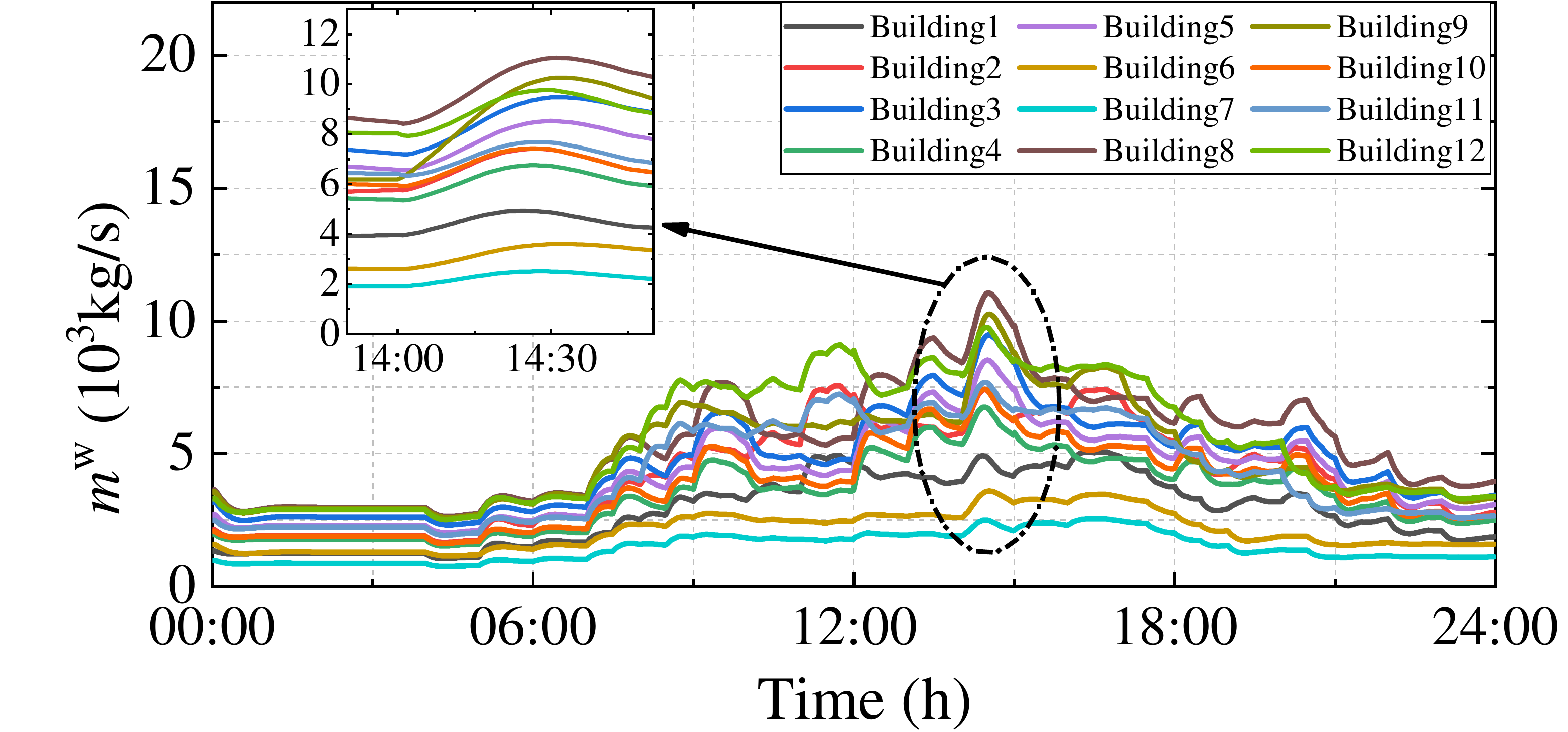}\label{case_physical_wind_a}}
	\subfigure[]{\includegraphics[width=0.88\columnwidth]{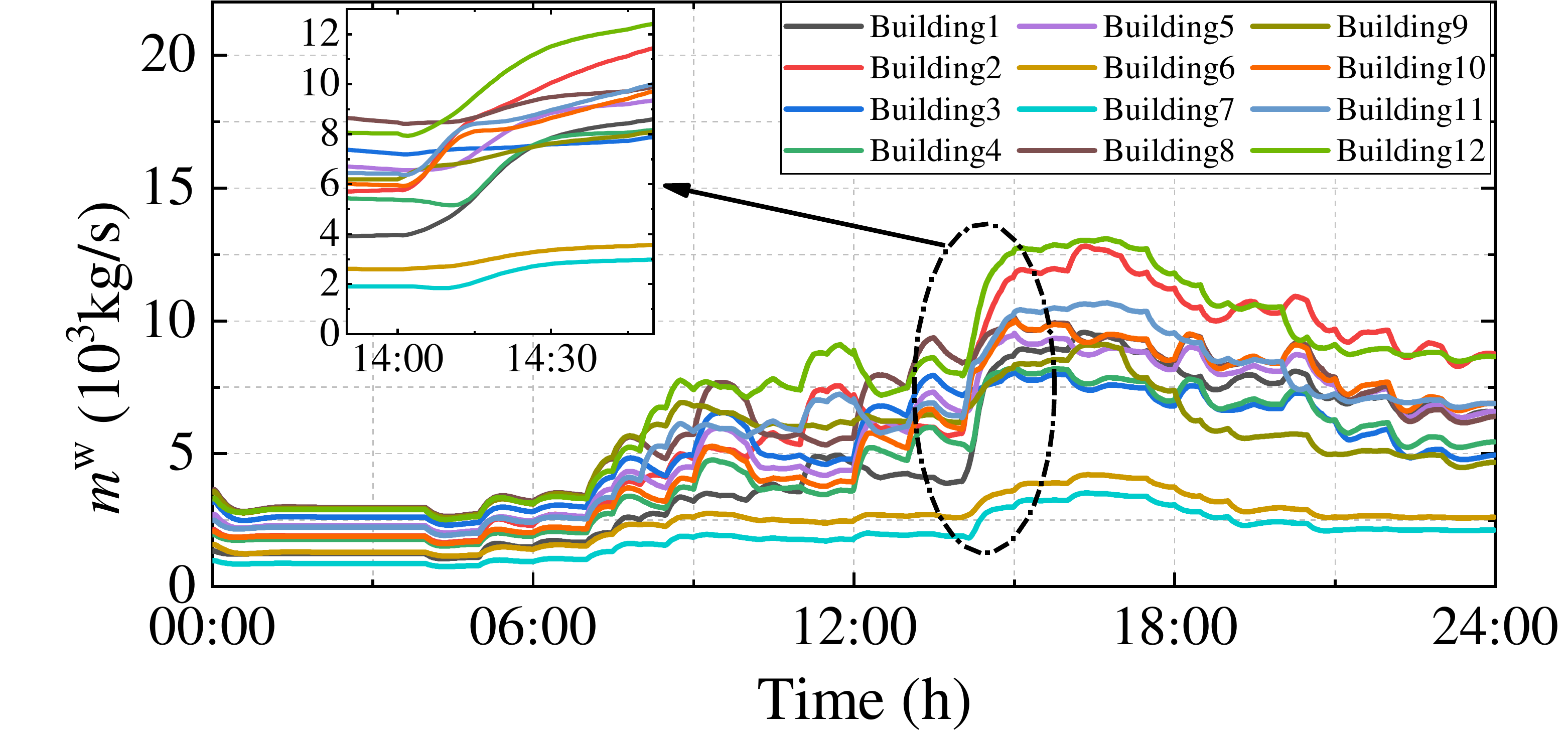}\label{case_physical_wind_b}}
	\caption{\textcolor{black}{The wind mass flow in buildings based on (a) PI controller; (b) Safe-DRL method.}}
	\label{case_physical_wind}
% 	\vspace{-4mm}
\end{figure}

\textcolor{black}{Buildings' wind mass flows $\bm{m}^\text{w}_t$ are dependent variables during the control process. If buildings' indoor temperature is higher than the set value, wind mass flows $\bm{m}^\text{w}_t$ will increase to cool down the buildings. The dynamic processes of the wind mass flow in different buildings are illustrated in Fig.~\ref{case_physical_wind}, where Fig.~\ref{case_physical_wind_a} and Fig.~\ref{case_physical_wind_b} are the control results of the PI controller and safe-DRL, respectively.
In the power reduction stage, wind mass flows in two figures both increase automatically due to the decrease of chilled water. The increase speed in Fig.~\ref{case_physical_wind_b} is faster than that in Fig.~\ref{case_physical_wind_a}, because a faster power decrease in the safe-DRL method makes a sharper increase of buildings' indoor temperatures. Buildings want to increase their cooling winds to cool down the indoor temperature.}

\textcolor{black}{In the power recovery stage, wind mass flows in Fig.~\ref{case_physical_wind_b} decrease more slowly than that in Fig.~\ref{case_physical_wind_a}. It means buildings' indoor temperatures are recovered more slowly using the safe-DRL method for preventing the power rebound. However, in the PI controller, the water mass flows are increased quickly in the recovery stage, which results in the quick decrease of wind mass flows in Fig.~\ref{case_physical_wind_a}.
To sum up, the proposed safe-DRL method can slow down the adjustment of wind mass flows to avoid the power rebound.}

\subsection{Sensitivity of the Key Parameters of DCS for Providing Operating Reserve}\label{sensitive}
\begin{figure}%[H]
	%\vspace{-2mm}
	\subfigbottomskip=-4pt
	\subfigcapskip=-4pt
	\centering
	\subfigure[]{\includegraphics[width=1\columnwidth]{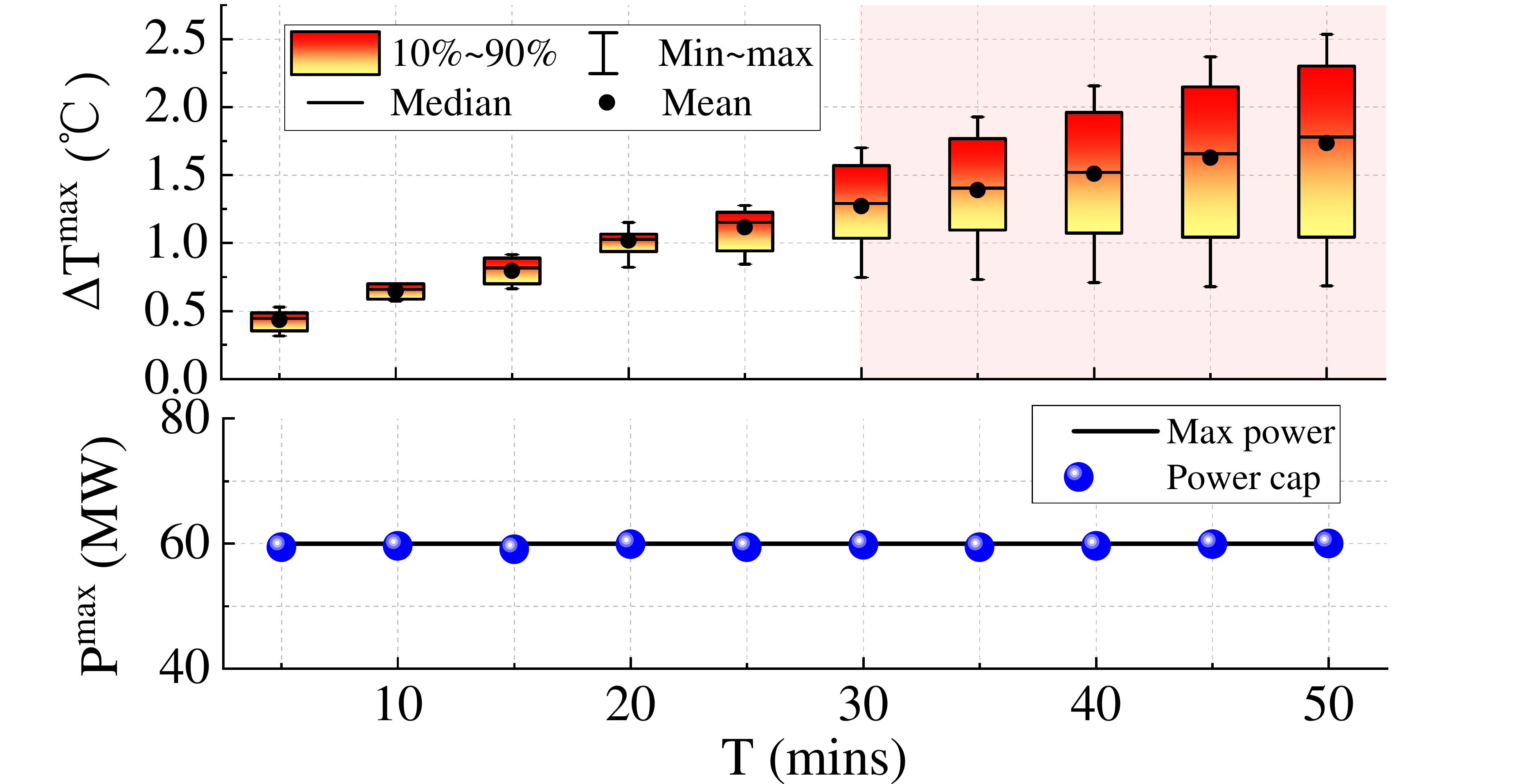}\label{case_sensitive_duration}}
	\subfigure[]{\includegraphics[width=1\columnwidth]{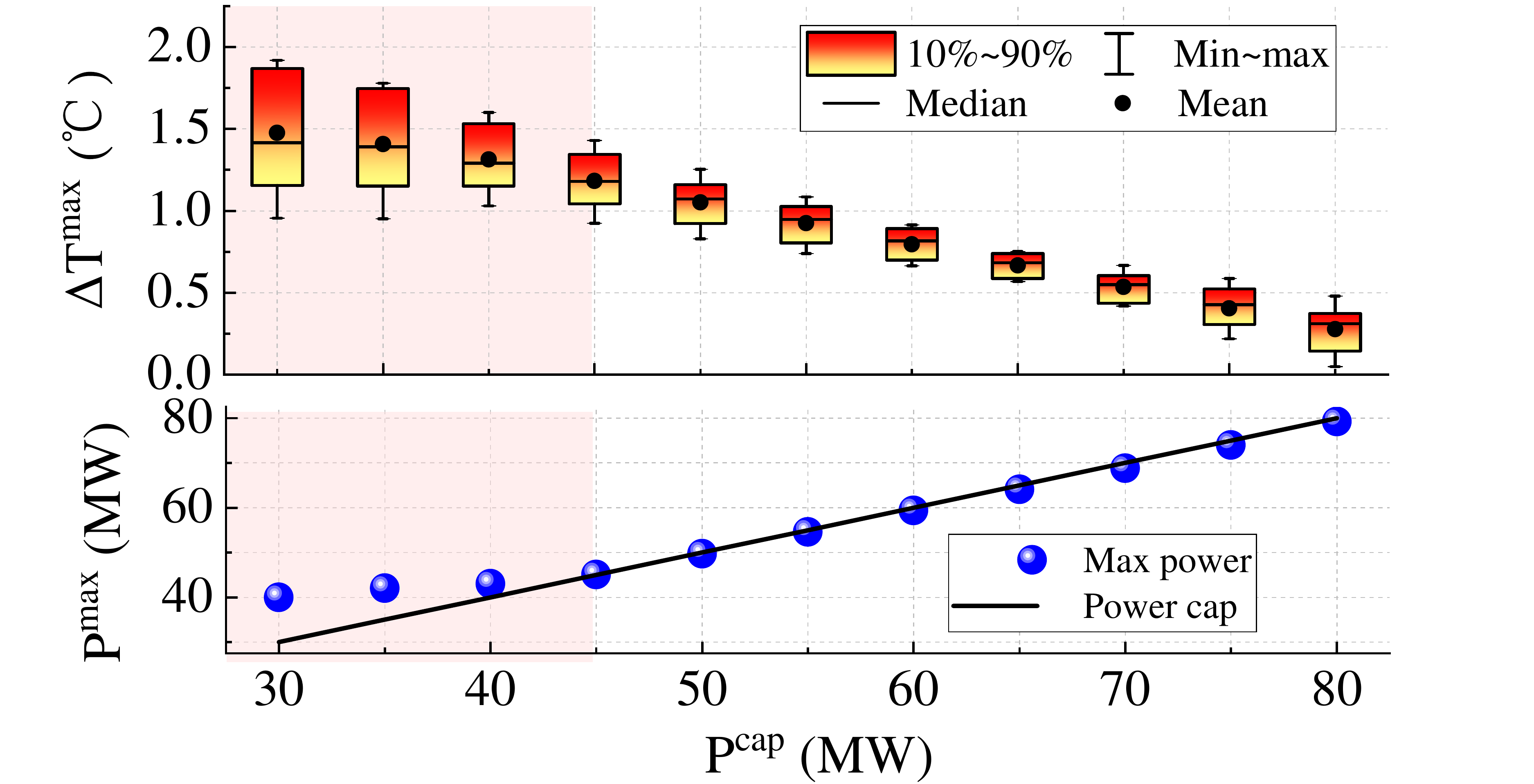}\label{case_sensitive_power_cap}}
	\caption{\textcolor{black}{The sensitivity results of (a) Duration; (b) Power cap.}}
	\label{case_sensitive}
	\vspace{-4mm}
\end{figure}

\textcolor{black}{Considering the real-time demands from power systems, the service requirements of the power cap and duration are probably various. Thus, the sensitive analysis is carried out to validate the effectiveness of the proposed method in different scenarios.
Fig.~\ref{case_sensitive_duration} and Fig.~\ref{case_sensitive_power_cap} show the analysis results of different duration periods and power caps, respectively. There are two observations: each building' maximum temperature deviation $\Delta T^\text{max}$, and the maximum power consumption $P^\text{max}$. The $\Delta T^\text{max}$ is for representing the indoor temperature comfort in buildings, and $P^\text{max}$ is for quantifying power cap violations.}

\textcolor{black}{Fig.\ref{case_sensitive_duration} shows the control results based on different scenarios of duration time, which ranges from 5 minutes to 50 minutes, where the power cap is set as 60MW.
It can be seen that the power cap can be always satisfied during the regulation period with different duration time. However, the impacts on buildings' indoor temperature are more significant with the increase of duration time. When the duration time is longer than 30 minutes, some buildings' temperature deviations can be up to 1.5$^\circ$C. By contrast, when the duration time is less than 30 minutes, all the buildings' temperature deviations are comparatively marginal with small variances. 
}

\textcolor{black}{Fig.\ref{case_sensitive_power_cap} shows the control results based on different required power caps, which range from 30MW to 80MW with the same duration time 15 minutes. It can be seen that the DCS is capable to provide high-quality operating reserve when the power cap is higher than 45 MW. However, the power cap can not be satisfied when it is lower than 45 MW, because DCS has the minimum physical limits in water mass flows. 
Besides, a lower power cap can result in larger temperature impacts on buildings because of less cooling supplies. It can be seen from Fig.\ref{case_sensitive_power_cap} that the average temperature deviation is out of the comfortable range and the corresponding variances are large, when the power cap is lower than 45 MW.
}

\textcolor{black}{In summary, the proposed safe-DRL can generally perform well with different service duration and power cap scenarios. However, long service duration time and low power cap may deteriorate its performance. A reasonable power cap and service duration time are significant for the final control results of DCS, which are generally determined by the capacity offer of DCS in electrical market\cite{market2018}. The strategy for the DCS operator to provide its operating reserve capacity is beyond the scope of this paper, but will be our future work.
}

\section{Conclusion}\label{dcs_conclusion}
This paper proposes a model-free safe-DRL scheme for DCS control problem to provide operating reserve. 
\textcolor{black}{A safe layer is proposed to effectively guarantee the critical power constraint in the power reduction stage. A self-adaptive target method is further adopted to tackle the power rebound in the power recovery stage. Meanwhile, it minimizes the impacts on buildings' indoor temperature to keep all the buildings as comfortable as possible.}
Numerical studies show that the DCS's operating power is always below the power cap during training, which ensures the ``safety" for providing operating reserve. Besides, the DCS's operating power can recover smoothly and avoid an undesirable peak power rebound. All the buildings' temperature deviations can be guaranteed within the required range $\pm$ 1$^\circ$C to stay comfortable.

\textcolor{black}{The well-trained agent in the proposed safe-DRL method can usually take effects on the similar DCS. If the physical system is totally different, the agent probably should be trained again using the new system's historical data. To be more efficient, the agent can transfer the old system's knowledge to the new system by less training episodes, which also will be our future work.}
\textcolor{black}{For the ancillary service that requires a long duration (e.g., more than one hour), it is probably more efficient for DCS to regulate both the chilled water's mass flow and supply temperature, which will be studied in our future work as well.}

\bibliographystyle{IEEEtran}
\bibliography{ref}

\end{document}